\expandafter\def\csname ver@fixltx2e.sty\endcsname{}
\documentclass{article}
\usepackage{geometry}
\usepackage{microtype}
\usepackage{graphicx}
\usepackage{subcaption}
\usepackage{booktabs} %
\usepackage{siunitx}
\usepackage{dblfloatfix}
\usepackage{caption}
\usepackage{subcaption}
\usepackage{hyperref}
\usepackage{graphicx}
\usepackage{subcaption}
\usepackage[dvipsnames]{xcolor}
\usepackage{authblk}
\usepackage{hyperref}

\usepackage{amsmath}
\usepackage{amssymb}
\usepackage{mathtools}
\usepackage{amsthm}
\usepackage[capitalize,noabbrev]{cleveref}
\usepackage{comment}
\usepackage[textsize=tiny]{todonotes}
\usepackage{authblk}

\usepackage{multirow}
\usepackage{xcolor}

\begin{document}
\onecolumn

\title{Functional Brain-to-Brain Transformation with No Shared Data}

\author[]{Navve Wasserman}
\author[]{Roman Beliy}
\author[]{Roy Urbach}
\author[]{Michal Irani}

\affil[]{The Weizmann Institute of Science}


\date{}

\maketitle

\vskip 0.3in

\begin{abstract}

Combining Functional MRI (fMRI) data across different subjects and datasets is crucial for many neuroscience tasks. Relying solely on shared anatomy for brain-to-brain mapping is inadequate. Existing functional transformation methods thus depend on shared stimuli across subjects and fMRI datasets, which are often unavailable. In this paper, we propose an approach for  computing functional brain-to-brain transformations {without any shared data}, a feat not previously achieved in functional transformations. This presents exciting research prospects for merging and enriching diverse datasets, even when they involve distinct stimuli that were collected using different fMRI machines of varying resolutions (e.g., 3-Tesla and 7-Tesla). Our approach combines brain-to-brain transformation with image-to-fMRI encoders, thus enabling to learn functional transformations on {visual} stimuli to which subjects were never  exposed. Furthermore, we demonstrate the applicability of our method for improving image-to-fMRI encoding of subjects scanned on  older low-resolution 3T fMRI datasets, by using a new high-resolution 7T fMRI dataset  (scanned on different subjects and different stimuli).


\end{abstract}

\section{Introduction}

\begin{figure*}
     \centering
     \begin{subfigure}{\textwidth}
         \centering
         \includegraphics[width=\textwidth]{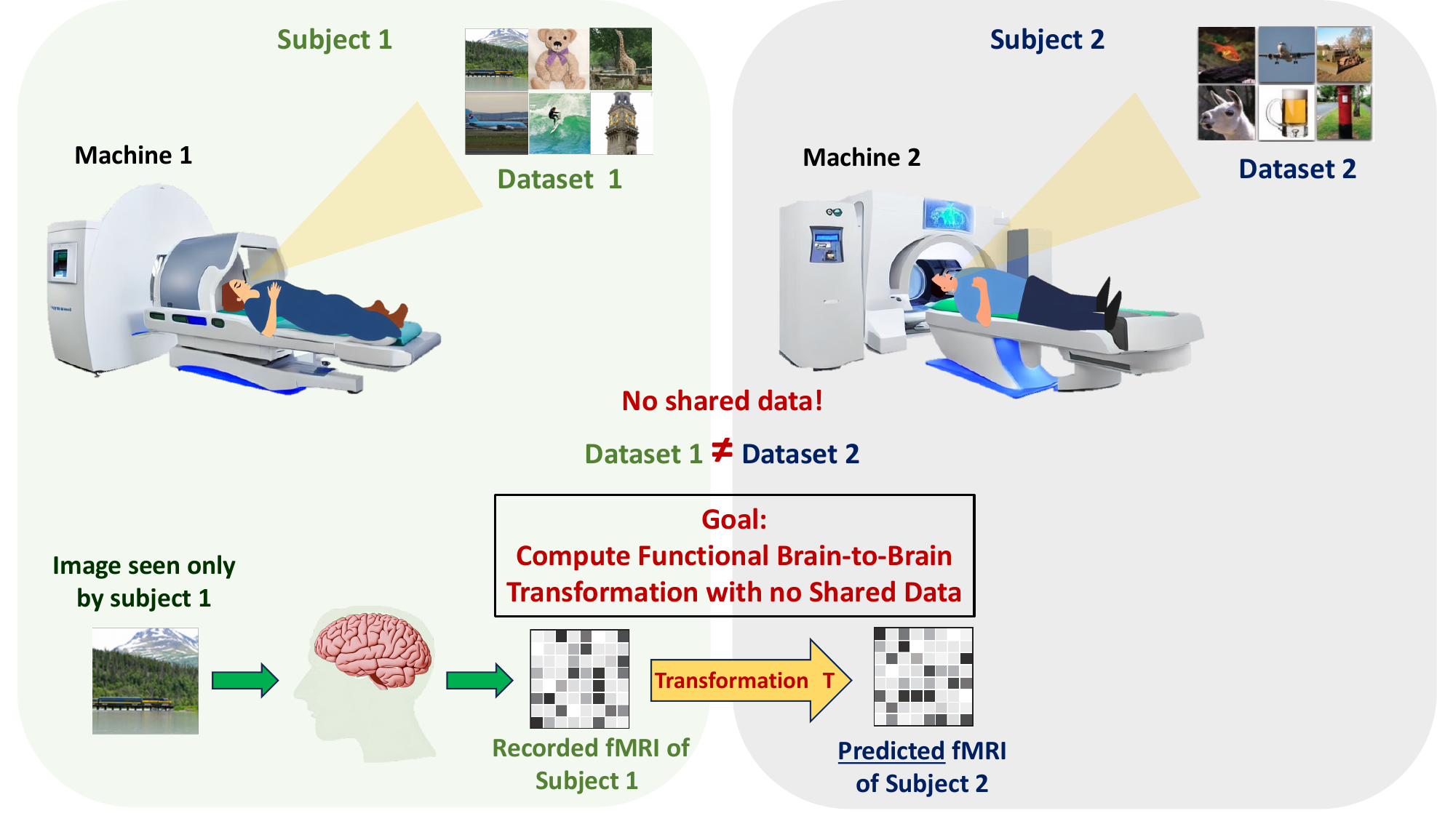}
     \end{subfigure}
 \caption{\textbf{Problem Formulation - Compute functional brain-to-brain transformation with no shared data}: This figure illustrates our task, learning brain-to-brain transformation between individuals that were exposed to distinct stimuli (images), and their fMRI scans were captured using different fMRI machines (possibly with different scanning resolutions, e.g., 7-Tesla and 3-Tesla). }
 \label{1}
\end{figure*}

Functional MRI (fMRI) has emerged as a powerful non-invasive method for measuring human brain activity
, allowing researchers to observe which areas of the brain are involved in different functions and behaviors~\cite{kanwisher1997fusiform,epstein1998cortical,downing2001cortical,tang2017using,heeger2002does}. However, the time-consuming and expensive data acquisition process of fMRI often leads to limited data availability. Moreover, inter-subject variability in brain structure and functionality poses additional challenges in analyzing fMRI data from multiple subjects, as different individuals may exhibit distinct brain activity patterns even in response to similar stimuli. To address these limitations, functional brain-to-brain transformation has been proposed. Functional brain-to-brain transformation involves learning inter-subject transformations that map the fMRI signals of one subject to the fMRI signals of another. Those transformations allow for the use of data collected from multiple subjects, thereby enriching the available data and enabling comparisons of brain activity across individuals and groups.

While the visual cortex exhibits similarities in topography and organization of functional regions across humans, there are fine-grained functional differences between individuals~\cite{riddle1995individual,frost2012measuring, conroy2013inter, zhen2015quantifying}. This variability poses challenges in comparing and mapping fMRI activation between individuals. As such, traditional anatomical alignment methods~\cite{mazziotta2001probabilistic,talairach19883,fischl2012freesurfer,dale1999cortical} have limited mapping prediction accuracy across subjects~\cite{mazziotta2001probabilistic,haxby2011common, yamada2015inter,brett2002problem}. To address this issue, functional alignment methods have been developed to match the functional behaviors of different individuals by learning relationships between brain activity patterns~\cite{haxby2011common, yamada2015inter, guntupalli2016model,chen2015reduced,de2010against,thual2022aligning,robinson2014msm}. There are two main approaches in functional alignment: the first maps subjects into a high-dimensional shared space, with the most well-known method being Hyper-alignment~\cite{haxby2011common,lorbert2012kernel,xu2012regularized,haxby2020hyperalignment}, and the second transforms directly between fMRI of two subjects using a conversion model~\cite{yamada2015inter}. \emph{\textbf{However, training these methods requires a sufficient amount of ``shared data" namely, fMRI obtained by showing the same stimuli to different subjects. This limits the number of available training examples, and makes it impossible to learn transformations across subjects which have no shared data.}}
The requirement for shared data thus undermines the ability to combine information from multiple different fMRI datasets, accumulated over the years from a wide variety of visual stimuli from different individuals.

In this paper, we introduce for the first time, an approach for computing functional brain-to-brain transformation without any shared data{, and show its effectiveness for fMRI responses to visual stimuli}. Figure~\ref{1} formulates the problem setting. Our approach (illustrated in Figure~\ref{2}), employs ``visual encoders", which are trained individually on subject-specific data, to encode images into subject-specific fMRI. Our method effectively uses visual encoders to predict fMRI of images not seen by a subject. This allows to use also \emph{\textbf{``non-shared data"}} (which refers to images seen by only one subject), and \emph{\textbf{``external data"}} (which refers to images not seen by any of the subjects, i.e. arbitrary images without fMRI recordings). To the best of our knowledge, we are the first to employ visual encoders to compute brain-to-brain transformations, thereby exploiting both ``non-shared" and ``external" data. By leveraging this approach, we significantly enrich the transformation training data {(i.e., the  number of examples used for learning the functional mapping between subjects)}. Notably, we demonstrate the capability to compute functional brain-to-brain transformations without any shared images, \emph{\textbf{even when the fMRI data was collected using different fMRI scanners}} (e.g., 7T vs 3T). Furthermore, our results show that we can improve one subject's image-to-fMRI encoder by utilizing another subject's encoder, even in the absence of shared data.  In particular, we demonstrate the ability to enhance the image-to-fMRI encoding of a subject trained on a low-resolution 3-Tesla fMRI dataset (e.g., ``GOD" dataset~\cite{Horikawa2017GenericFeatures}) by incorporating another subject's encoder trained on a new 7-Tesla fMRI dataset (e.g., NSD~\cite{allen2022massive}) using our proposed method.

\vspace*{0.2cm}
\noindent
Our contributions are therefore:
\begin{itemize}
        \item Introducing functional brain-to-brain transformation without the need for shared data, even across distinctly different image datasets and across fMRI scanners of different resolutions (e.g., 7T vs 3T).
        \item Enhancing the encoding of one subject using {higher quality} data from another subject (both within and across datasets, e.g., utilizing a subject from a new 7-Tesla dataset to enhance a subject from a low-resolution 3-Tesla dataset).
\end{itemize}

\vspace{-0.35cm}
\section{Overview of the Approach}
\vspace{-0.05cm}
Our approach aims to learn Brain-to-Brain {(B2B)} transformations between subjects (within and across datasets), when the subjects were exposed to different stimuli, possibly acquired by different fMRI machines (see Figure~\ref{1}). We first explain how using visual encoders allows generating ``corresponding" fMRI data between different subjects without any shared stimuli. {By ``corresponding" fMRI we mean synthesizing the fMRI responses of two different subjects, as if they were exposed to the same image stimulus.} This ``corresponding" fMRI data is used to train the {B2B-}transformation{ across different subjects}. We further demonstrate the effectiveness and applicability of our {B2B-transformations} for improving the {Image-to-fMRI} encoding of one subject using {higher-quality fMRI data} from another subject (scanned on a different stimuli dataset and on a different fMRI scanner). \\
\\
\subsection{Brain-to-Brain (B2B) transformation with no shared data}
\begin{figure*}
     \centering
     \begin{subfigure}{\textwidth}
         \centering
         \includegraphics[width=\textwidth]{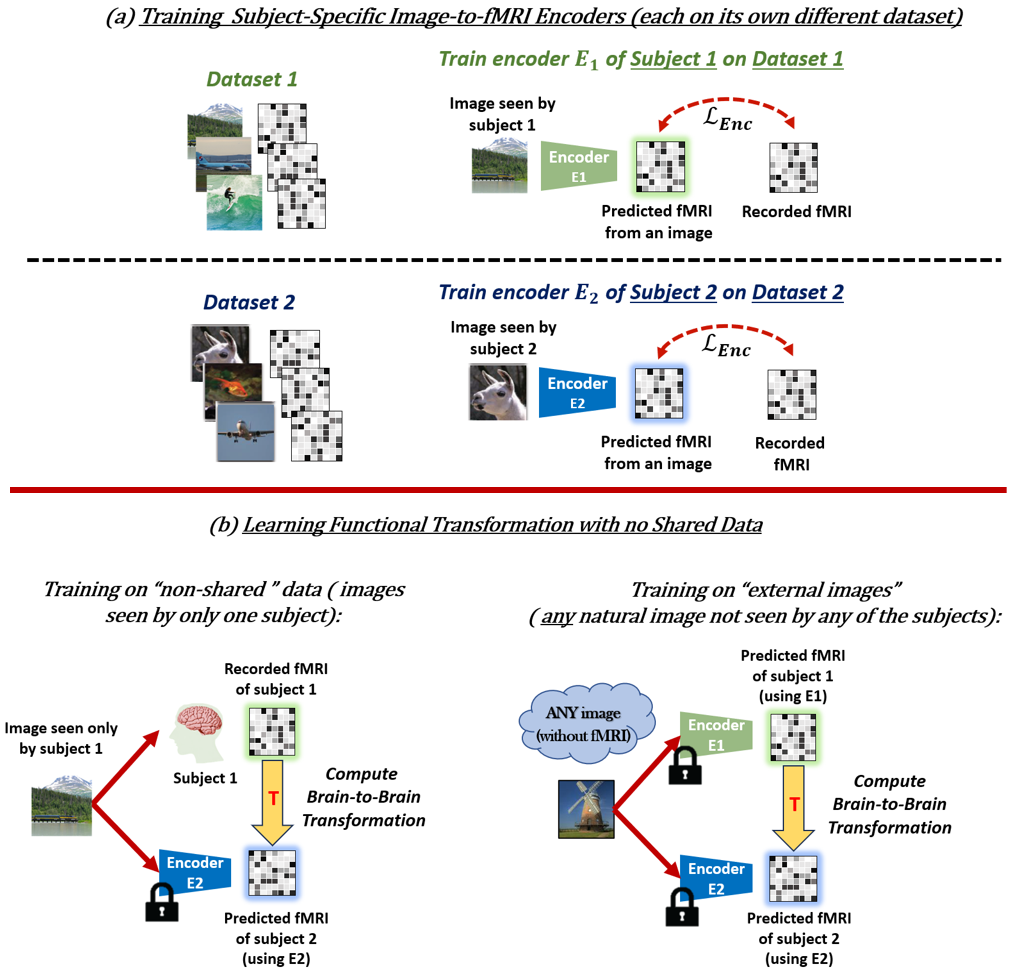}
     \end{subfigure}
 \caption{\textbf{An overview of the approach - Estimating functional brain-to-brain transformations with no shared data}:
 \textbf{(a)} {We first independently train a visual encoder} for each subject using its own {dataset of} image-fMRI pairs. Each subject's data can potentially come from entirely different image datasets and different fMRI scanners.
\textbf{(b)} {We employ the subject-specific pre-trained encoders to predict fMRI responses both for ``non-shared" images (i.e, images seen by only one subject), and for ``external" images (any natural image never seen by any of the subjects). The transformation $T$ is then computed by training a single linear layer to correctly transform fMRI patterns of Subject1 to those of Subject2 using the large set of predicted fMRIs. The lock symbol signifies that the encoder weights are fixed and not updated while training the transformation (in this scenario).}}
 \label{2}
\end{figure*}

Our approach involves the use of ``visual encoders", which encode images into fMRI signals. These are modeled via deep neural networks, trained on subject-specific image-fMRI pairs. Image-to-fMRI encoders were shown to obtain impressive accuracy in forecasting a subject's fMRI responses to novel images~\cite{yamins2014performance, eickenberg2017seeing, wen2018neural, wen2018deep, beliy2019voxels, gaziv2022self}. We first individually train a visual encoder for each subject, utilizing their own respective image-fMRI data (see Figure~\ref{2}.a). Note that such encoder training does \textbf{not} require for the two subjects to have experienced the same stimuli; they can belong to entirely distinct image-fMRI datasets.

The functional brain-to-brain transformation $T$ is trained to map the fMRI patterns of Subject1 to their corresponding counterparts in Subject2. The straightforward approach to train the transformation network is to transform an fMRI scan from one subject to match the corresponding fMRI scan of the second subject, and then evaluate its error w.r.t the recorded fMRI (e.g., using the mean square error loss). However, this approach relies on the presence of shared data, and is moot when such shared data is unavailable. Leveraging the predictive capability of the pre-trained encoders for predicting fMRI responses of images, we can generate corresponding fMRI patterns for ``non-shared" and ``external" images. {For example, given a ``non-shared'' image $\mathcal{I}_{Ns_1}$ seen only by Subject1, and Subject1's fMRI scan on that image, 
$\mathcal{F}_{1}(\mathcal{I}_{Ns_1})$}
 (as shown in Figure~\ref{2}.b on the left side), we can approximate the corresponding fMRI pattern for Subject2 by inputting image~{$\mathcal{I}_{Ns_1}$} into the encoder~${E}_{2}$ designed for Subject2. We then optimize the learned transformation T {(modeled by a single linear layer)} using the fMRI loss between the transformed fMRI~{$T(\mathcal{F}_{1}(\mathcal{I}_{Ns_1}))$} and the encoded fMRI~{${E_2(\mathcal{I}_{Ns_1})}$}, aiming to minimize differences between the two predicted fMRIs: {($||T(\mathcal{F}_{1}(\mathcal{I}_{Ns_1})) - E_2(\mathcal{I}_{Ns_1})||$)}. A similar process can be carried out using the fMRI {scan of a ``non-shared''} image observed solely by Subject2, {$\mathcal{F}_{2}(\mathcal{I}_{Ns_2})$}, comparing it against {the transformed fMRI} encoding of the same image using the encoder of Subject1: 
{($||T({E}_{1}(\mathcal{I}_{Ns_2})) - \mathcal{F}_{2}(\mathcal{I}_{Ns_2}))||$)}. Furthermore, we can incorporate ``external" images (depicted in Figure~\ref{2}.b on the right side). These images are natural images that were never seen by any of the subjects, and thus do not have any fMRI recordings. We feed these ``external" images into the encoders of both subjects, subsequently transforming the encoded fMRI pattern of Subject1 and comparing it with the encoded fMRI pattern of Subject2: {($||T({E}_{1}(\mathcal{I}_{Ext})) - E_2(\mathcal{I}_{Ext})||$)}. This provides ``infinitely" many training data for \emph{training} of the brain-to-brain transformation $T$. 
{While \emph{real} fMRI scans are generally more reliable than \emph{encoded} fMRIs predicted from images, the primary advantage of using external images lies in their infinite supply, which is particularly beneficial when the number of available fMRI-image pairs is small.}\\

\subsection{Improving Image-to-fMRI Encoding via B2B Transformation} \ \
To showcase the power of our method, we employ it to enhance the image-to-fMRI encoder of one subject using another subject with superior data quality or more examples. For simplicity, we term this technique the ``teacher-student" method, where the subject with higher quality data plays the role of the ``teacher", and the poorer-quality subject is the ``student". This ``teacher-student" approach is applicable to subjects of different quality within the same dataset, and more interestingly, it can be employed across different datasets and machines using the above-described method.

We start by training the ``teacher" subject's encoder ($E_{t}$) using all its available data. Once trained, we keep its weights unchanged. Then, we simultaneously train the ``student" subject's encoder ($E_{s}$) and the transformation $T$ that maps the fMRI data between the ``teacher" and ``student" subjects ($T_{t->s}$). This joint training benefits both the encoder and the transformation, allowing them to improve together. During this step, the ``student" encoder is not only trained using its own image-fMRI data, it also incorporates information from the ``teacher" encoder and its unique fMRI data (exploit ``non-shared" and ``external" data). This process is similar to the one shown in Figure~\ref{2}.b, but with a difference: now, the ``student" encoder can be adjusted during training.

For instance, let's say we take the actual fMRI response from the ``teacher" corresponding to an image that the ``student" has not encountered before. This response can be transformed to the ``student" fMRI space and compared with the encoded response produced by the ``student" encoder given that image. Moreover, to make use of ``external" data, we can take any image, predict its fMRI using the ``teacher" encoder, transform that prediction to the ``student" fMRI space, and then compare it with the encoded response of the same ``external" image produced by the ``student" encoder.

\vspace{-0.2cm}
\newpage
\section{Results}
\vspace{-0.1cm}

\begin{figure*}
 \centering
 \begin{subfigure}{\textwidth}
     \centering
     \includegraphics[width=\textwidth]{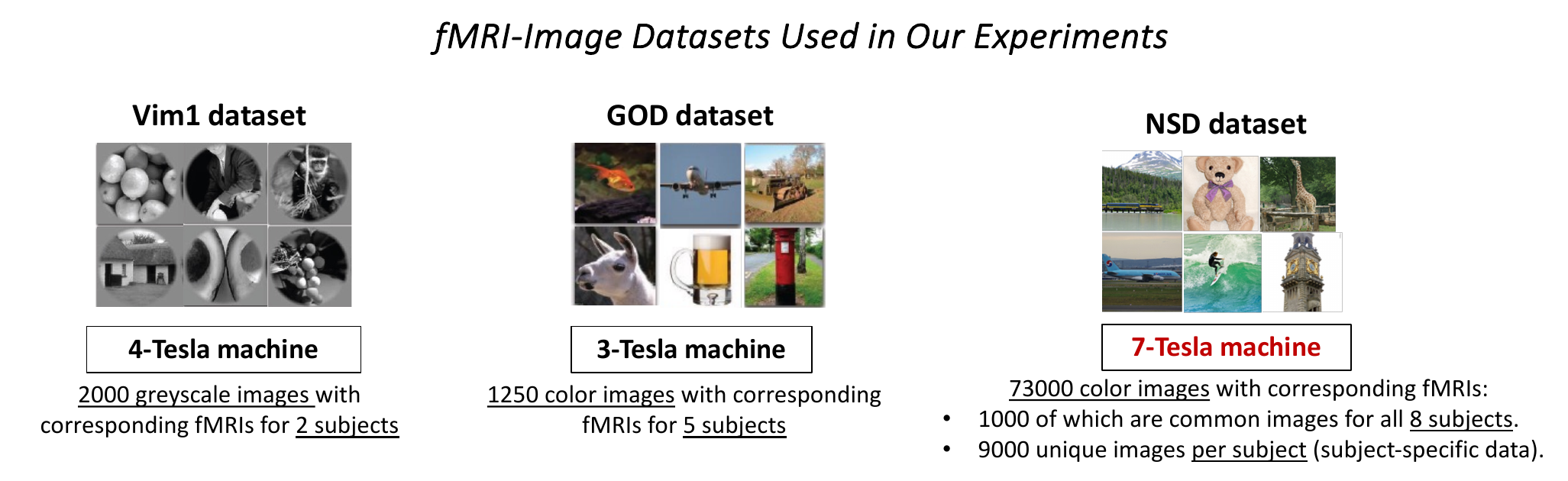}
     \label{Figure:Datasets}
 \end{subfigure}
\caption{\textbf{The fMRI datasets used in our experiments:}
(i)~``Vim-1" dataset~\cite{kay2008identifying} which features grey-scale images and their corresponding 4-Tesla fMRI recordings. (ii)~the ``Generic Object Decoding" (GOD) dataset~\cite{shen2019deep} which comprises natural ImageNet images with their corresponding 3-Tesla fMRI recordings. (iii)~the ``Natural Scenes Dataset" (NSD), a new 7-Tesla dataset~\cite{allen2022massive} with 8 subjects, each of whom viewed 1000 shared images and 9000 unique subject-specific images. {This resulted in a total of 73,000 different natural images (taken from the COCO dataset).}}
\label{3}
\end{figure*}

We have experimented with three prominent fMRI datasets (see Figure~\ref{3}): (i)~the ``vim-1" dataset~\cite{kay2008identifying,naselaris2009bayesian,kay2011fmri}, which {contains} around 2000 grey-scale images {(as opposed to color images)} and their corresponding 4-Tesla fMRI recordings for 2 subjects; (ii)~the ``Generic Object Decoding" (GOD) dataset~\cite{shen2019deep,horikawa2020attentionally,ho2022inter}, which comprises of 1300 pairs of natural images from ImageNet with 3-Tesla fMRI recordings for 5 subjects; and (iii)~the ``Natural Scenes Dataset" (NSD){~\cite{allen2022massive}, a new 7-Tesla dataset with 8 subjects, each of whom viewed 9000 unique \emph{subject-specific images},  in addition to $\sim$1000 \emph{shared images} viewed by all subjects. This resulted in a total of 73,000 different natural images (taken from the COCO dataset).}
{See \ref{Methods:Datasets} for details about train/val/test splits for each dataset.}

\subsection*{Results of Brain-to-Brain (B2B) Transformation with No Shared Data}

This section presents results which substantiate two central claims: (i)~Functional B2B transformations can be computed without relying on shared data, yielding significantly better results than anatomical mapping; (ii)~Harnessing a large number of ``non-shared" and/or ``external" images (either alongside or instead of shared data), leads to improved B2B transformations compared to those estimated  {when using only the small number of available} shared data. 

To validate these claims, we used the NSD dataset. We automatically selected 10,000 voxels that are most sensitive to visual tasks, based on their signal-to-noise ratio (SNR), which is a standard approach in the field (see Methods Section \ref{Sec:methods_data} for more details). This dataset offers a relatively limited number of shared data (1000 shared images) and a substantial number of non-shared data (9000 {\emph{subject-specific}} images per subject), rendering it well-suited for illustrating our claims. We train a transformation model {exclusively on the} \emph{non-shared data}, and then assess its performance on a dedicated evaluation set of \emph{shared data}. 
{For each subject we independently trained an Image-to-fMRI encoder on its own subject-specific image-fMRI pairs (see \cref{table:mean_correlations} for each encoder's prediction correlation). These encoders are then used for training the B2B transformations.}
We evaluate the learned B2B transformation $T$  {both quantitatively and qualitatively.}
\\

\begin{figure*}     
     \begin{subfigure}{0.95\textwidth}
        \centering
        \includegraphics[width=\textwidth]{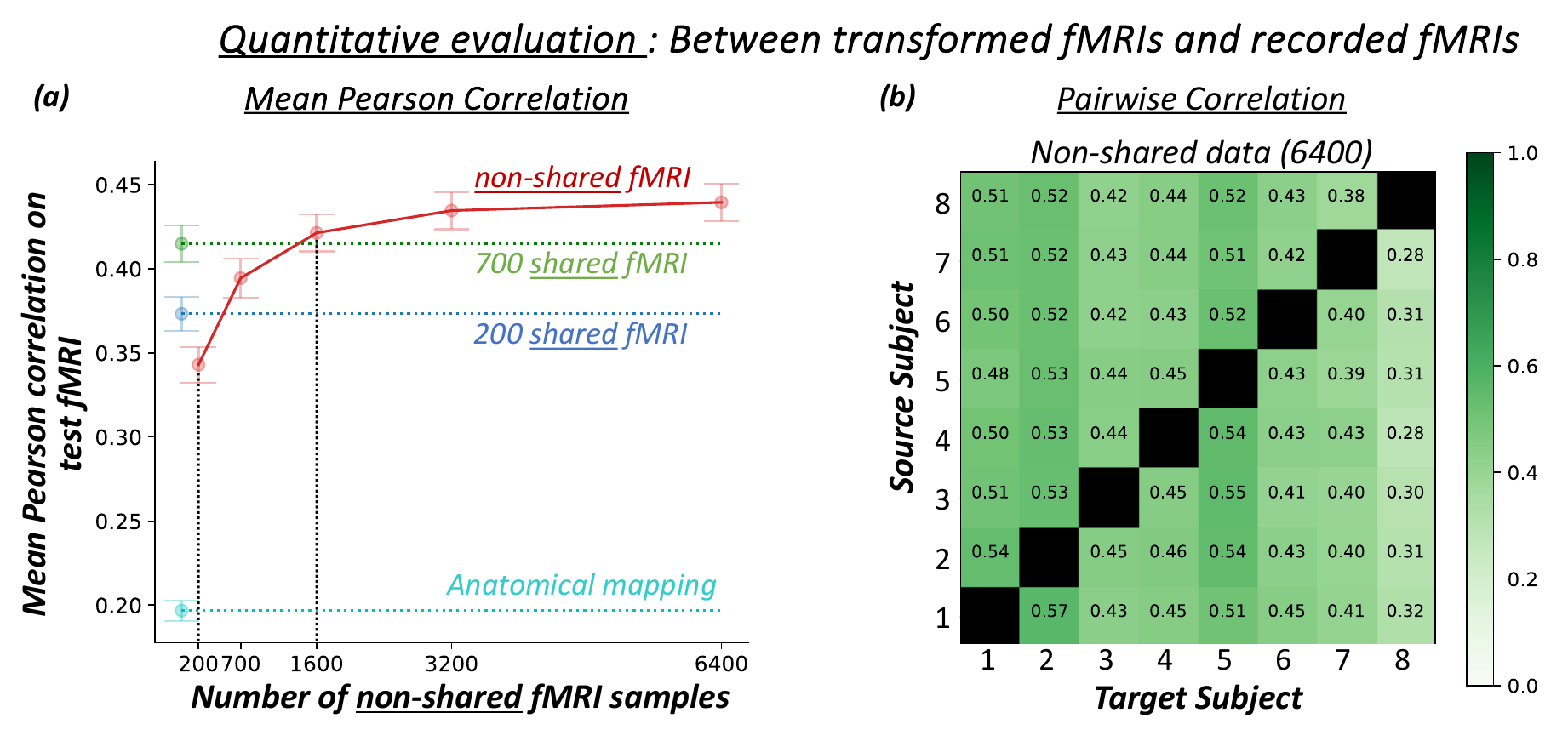} 
     \end{subfigure}
      \hfill
     \caption{\textbf{Quantitative evaluation of brain-to-brain transformation with no shared data}:
     \textbf{(a)}~ The graph presents the mean Pearson's r correlation averaged over all 56 transformations of all possible pairs of subjects in the NSD dataset, with SEM (standard error of the mean) error bars. We compare the resulting transformations when trained only with non-shared data (solid red line) to three other transformations: an anatomical mapping to a common brain space (cyan dotted line), a transformation trained with only 200 shared examples (dotted blue line), and a transformation trained with 700 shared examples (dotted grey line). The remaining 300 ``shared data" serve as ``test data" for assessing and comparing the quality of all the learned transformations. The x-axis of the graph represents the number of non-shared examples available for training, where our transformation model is the only one that can use non-shared examples.
     \textbf{(b)}~{The figure presents a correlation heatmap for all NSD subject pairs, using solely non-shared and external data.}
     }
    \label{Figure:Transformation_NSD}

\end{figure*}

\begin{figure*}     
    \centering
     \begin{subfigure}{0.92\textwidth}
         \includegraphics[width=\textwidth]{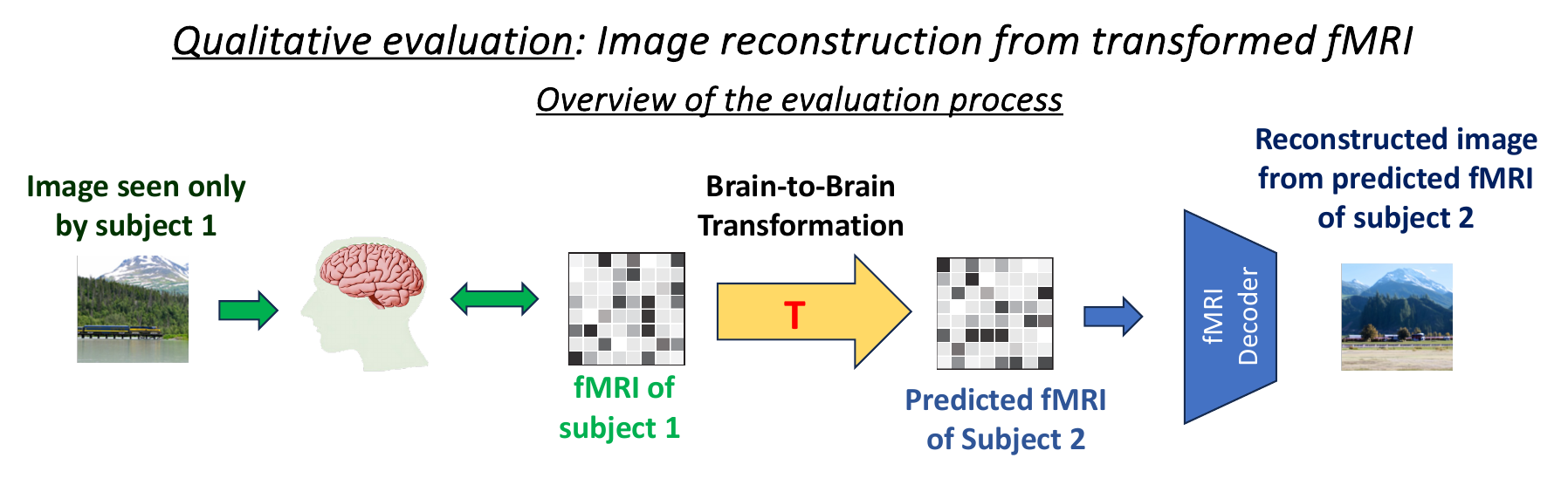} 
     \end{subfigure}
      \hfill
     \begin{subfigure}{0.92\textwidth}
         \includegraphics[width=\textwidth]{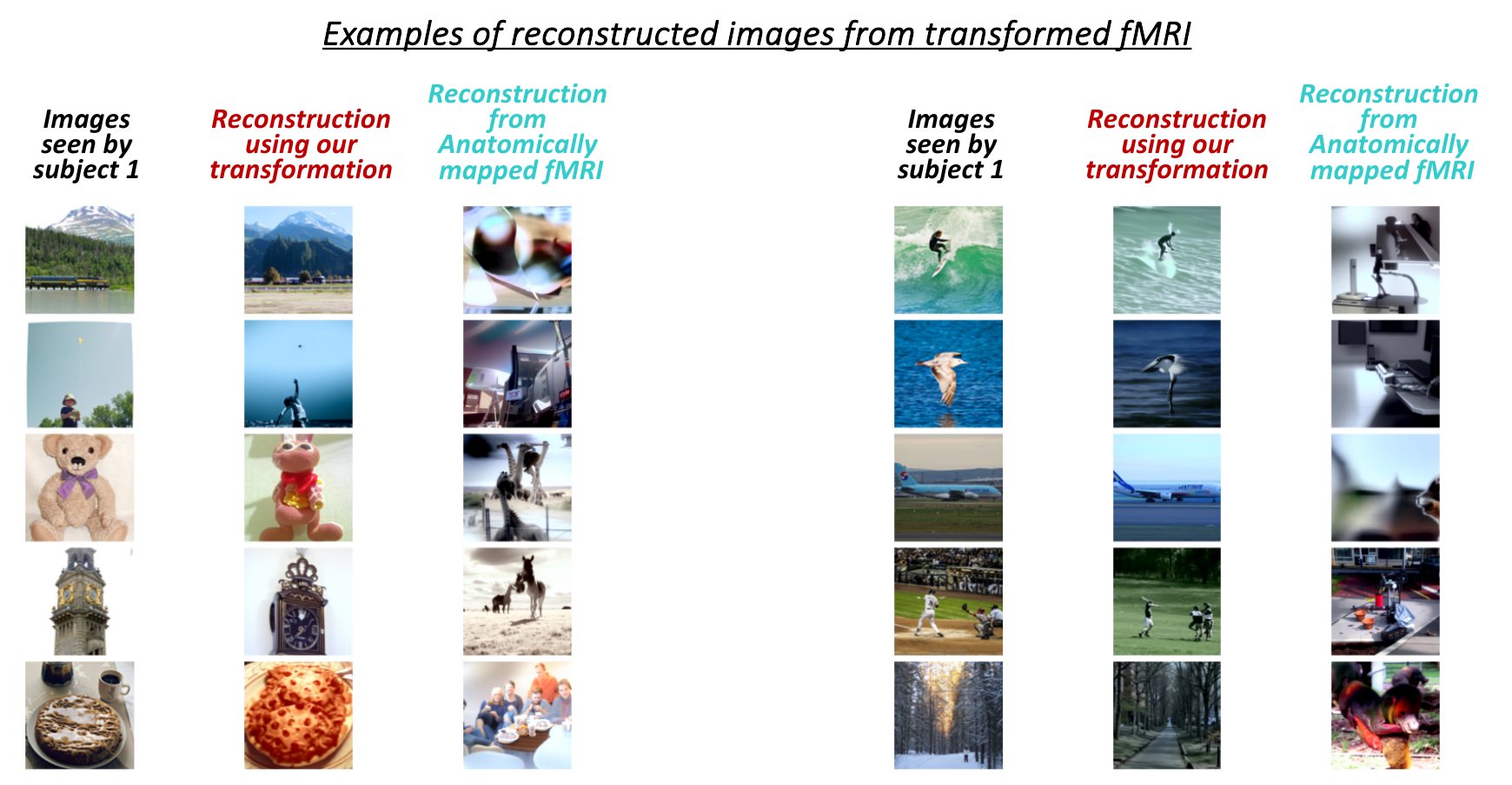} 
     \end{subfigure}
     \hfill
     \caption{\textbf{Qualitative evaluation of brain-to-brain transformation with no shared data}:
     This figure provides a visual comparison of image decoding when using our transformation model versus anatomical alignment method (the only 2 possible methods for handling non-shared data). As can be seen in the diagram, the transformation is used to map the fMRI activity of Subject1 to predict the fMRI activity of Subject2. Then we utilize the pre-trained decoder~\cite{gaziv2022self} of Subject2 (trained on his mutually exclusive data), together with a novel diffusion model decoding approach~\cite{scotti2023reconstructing} to generate images from the transformed fMRI. The figure displays the original image seen by Subject1, along with two reconstructed images post-fMRI transformation from Subject1 to Subject2: Once using anatomical mapping between the subjects, and once using our transformation model $T$ trained without any shared data.
    }
    \label{Figure:Transformation_Decoding}
\end{figure*}

\noindent
\textbf{\underline{Quantitative evaluation:}} We measure the Pearson correlation between the \emph{predicted} fMRI (i.e., the fMRI obtained by transforming the source subject's fMRI to the target subject's space), and the \emph{actual} fMRI of the target subject. The graph in \cref{Figure:Transformation_NSD}.a  presents the mean Pearson's r correlation averaged over all 56 transformations of all possible pairs of subjects in the NSD dataset. We compare the resulting transformations when trained with non-shared data {together with 6000 external images} (solid red line) to three other transformations: an anatomical mapping to a common brain space (cyan dotted line), a transformation trained with only 200 shared examples (dotted blue line), and a transformation trained with 700 shared examples (dotted grey line). The remaining 300 ``shared data" serve as ``test data" for assessing and comparing the quality of all the learned B2B transformations. The x-axis of the graph represents the number of non-shared examples available for training, where our transformation model is the only one that can use non-shared examples. When there are no shared examples whatsoever, we can only compare our method to anatomical alignment. As seen from the graph, our approach performs significantly better than anatomical alignment, even with a small number of non-shared examples (paired samples permutation test with \(N=56\), \(p<0.00001\), \(d=2.48\)) with up to 75\% improvement when training with as little as 200 non-shared examples). Furthermore, our method, when utilizing only non-shared examples, yields comparable or even better results than transformation models trained with shared data. As we increase the number of non-shared examples, the transformation model's performance continues to improve, surpassing models trained with only 200 and 700 shared examples. Although shared data is more beneficial when there is a low number of examples, 700 non-shared examples (M=0.407, SD=0.07) perform better than 200 shared examples (M=0.385, SD=0.07, \(N=56\), \(p<0.00001\), \(d=1.61\)), and 1600 non-shared examples (M=0.4349, SD=0.07) perform better than all 700 shared examples (M=0.429, SD=0.07, \(N=56\), \(p<0.001\), \(d=0.44\)). This is {likely} because, with sufficient examples, the encoder used for our training of the transformation is robust enough to compensate for the absence of shared examples. 

{In \cref{Figure:Transformation_NSD}.b we present a correlation heatmap of all NSD subject pairs trained using only non-shared and external data.} {Additional results in the Supplementary Material show transformations for specific subjects (\cref{SM_Figure:Tranformation_singles,SM_Figure:Transformation_hetmaps_NSD}), and demonstrating that combining non-shared examples with shared examples yields better performance than using shared examples alone (\cref{S2,S3}).} {Furthermore, the capability to learn B2B transformations without shared data enables us to train transformations across different datasets with differing fMRI resolutions. In the Supplementary Material (\cref{SM_Figure:Transformation_across_datasets}), we provide more detailed quantitative results of these B2B transformations (for any pair of subjects), to showcase this ability. This capability is later utilized to enhance the encoding of one dataset using another.}
 \\

\vspace*{0.3cm}
\noindent
\textbf{\underline{Qualitative evaluation:}} 
We further assess the quality of the 
B2B-transformed fMRI through an {fMRI-to-Image \emph{decoder}}, and \emph{visually} inspect the quality of the decoded images. This decoding is achieved by combining 2 fMRI-to-Image decoding methods -- \cite{gaziv2022self} and~\cite{scotti2023reconstructing} (for more details, please refer to {\cref{sec:method_decoding}}). \cref{Figure:Transformation_Decoding}.b  provides a qualitative visual comparison of {images \emph{decoded} from the transformed fMRI -- once after} using our B2B transformation model{, and once after anatomically mapping the fMRI} (the only two possible methods for handling non-shared data). The transformation is used to map the fMRI activity of NSD Subject1 to predict the fMRI activity of NSD Subject2. We then utilize {the decoder of Subject2 (pre-trained on its unique subject-specific data)}, to generate images from the transformed fMRI. The figure displays the original image seen by Subject1, along with two reconstructed images post fMRI transformation from Subject1 to Subject2: One obtained by using anatomical mapping between the two subjects, and one using our B2B-transformation model $T$ trained without any shared data. This demonstrates that our B2B-transformation produces fMRI predictions with more relevant and semantically meaningful information for reconstructing images. This visual comparison underscores that our transformation, trained without shared data, generates significantly superior transformed fMRIs compared to anatomical mapping (the only available option to-date). For additional reconstructions and transformations, please refer to Figures~\ref{S4} and~\ref{S5} in the Supplementary Material.

\subsection*{{Improving Image-to-fMRI Encoding via B2B Transformation}}

{Ours B2B transformation} allows one subject from a high quality fMRI dataset (the ``teacher") to improve the image-to-fMRI encoding capabilities of another subject from a low quality fMRI dataset (the ``student"). Normally there are no shared images across different fMRI datasets, in which case our B2B transformation method is needed. This applies also when both subjects are from the same dataset, but one subject has higher fMRI quality than the other. 
{We assess the significance and utility of the enhanced encoder through 3 evaluations: (i)~\emph{Pearson correlation} between predicted and measured fMRI; (ii)~the ability to perform accurate \emph{Image Retrieval} via the encoded fMRI; and (iii)~the influence of the encoded fMRI on \emph{Image Classification} accuracy.}

\begin{figure*}
     \centering
     \begin{subfigure}{\textwidth}
         \centering
         \includegraphics[width=\textwidth]{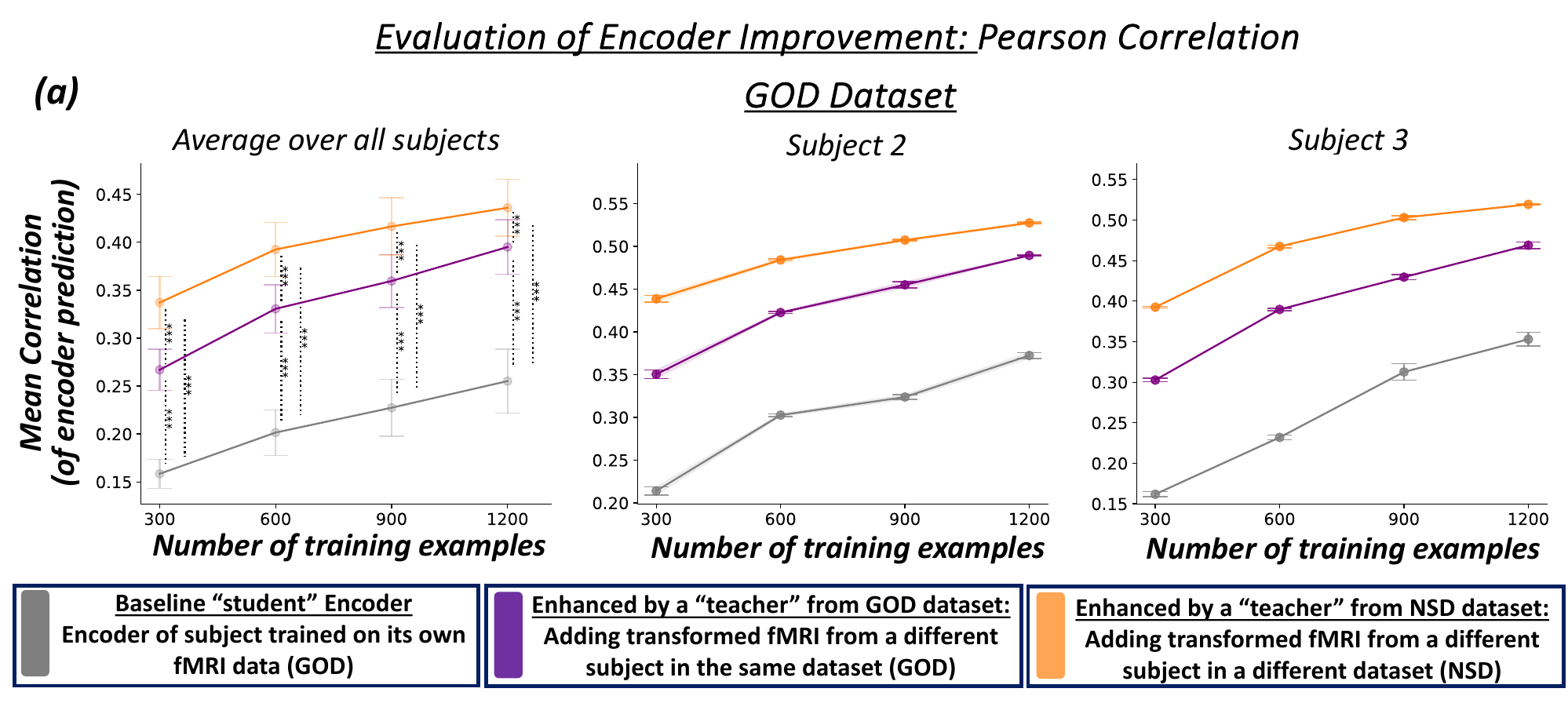}
     \end{subfigure}
     \begin{subfigure}{\textwidth}
         \centering
         \includegraphics[width=\textwidth]{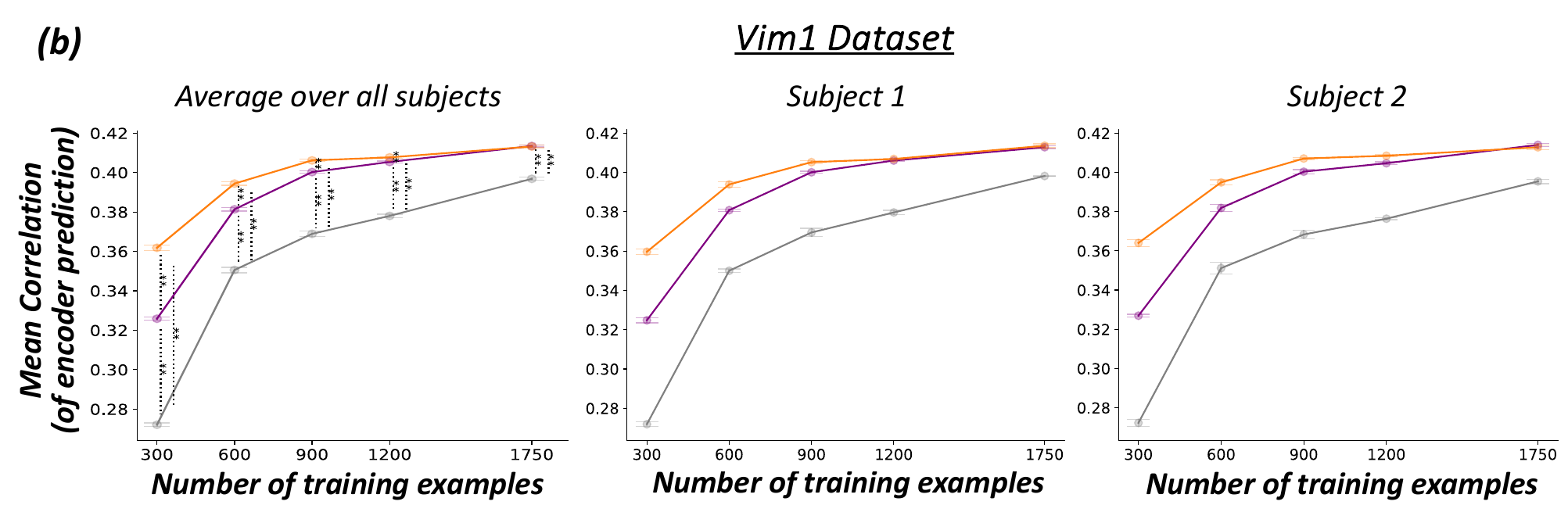}
     \end{subfigure}
 \caption{\textbf{Improving Image-to-fMRI Encoding via B2B Transformation}:
 This figure compares the quality of the ``student" baseline encoder (which is trained solely on the student's own fMRI data) to the student encoder obtained when incorporating also data from the ``teacher" {via our B2B transformation)}. In each plot, the grey line represents the student baseline encoder model, the purple line represents the encoder improvement obtained when using a ``teacher" from the GOD dataset (e.g., Subject4 is the highest-quality subject in the GOD dataset). The orange line corresponds to the encoder improvement achieved by utilizing a superior subject from another dataset (NSD), which has more scanned examples and higher fMRI resolution (7-Tesla machine). The quality of the resulting encoders is assessed through the mean Pearson correlation of all predicted fMRI voxels. The ``number of training examples" refers to the number of the student's own image-fMRI pairs of examples used for training the ``student" encoder, whereas the ``teacher" subject (drawn from either GOD or NSD dataset) has been trained using all its own available examples and remains unchanged during the training of the student encoder. The dotted lines mark which two types of models were statistically compared, and the number of asterisks (*) mark the significance (see Methods).
 \textbf{(a)} Mean Pearson correlation results averaged over all five subjects in the GOD dataset, as well as individual subject results.
\textbf{(b)} Mean Pearson correlation results averaged across the two subjects in the Vim1 dataset, as well as individual subject results.
}
\label{Figure:Encoder_GOD}
\end{figure*}

\begin{figure*}
     \centering
     \begin{subfigure}{\textwidth}
         \centering
         \includegraphics[width=\textwidth]{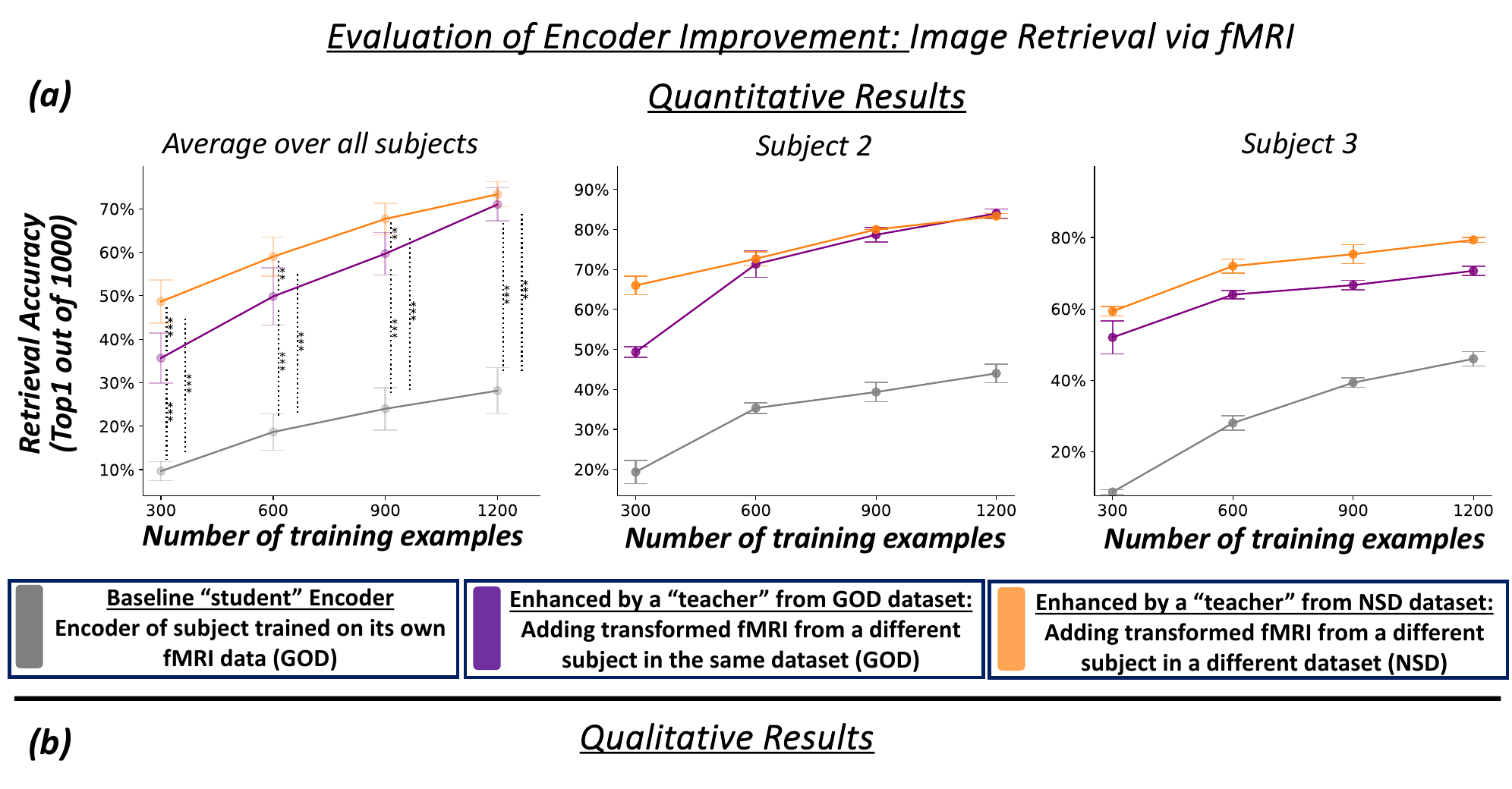}
     \end{subfigure}

     \begin{subfigure}{0.95\textwidth}
         \centering
         \includegraphics[width=\textwidth]{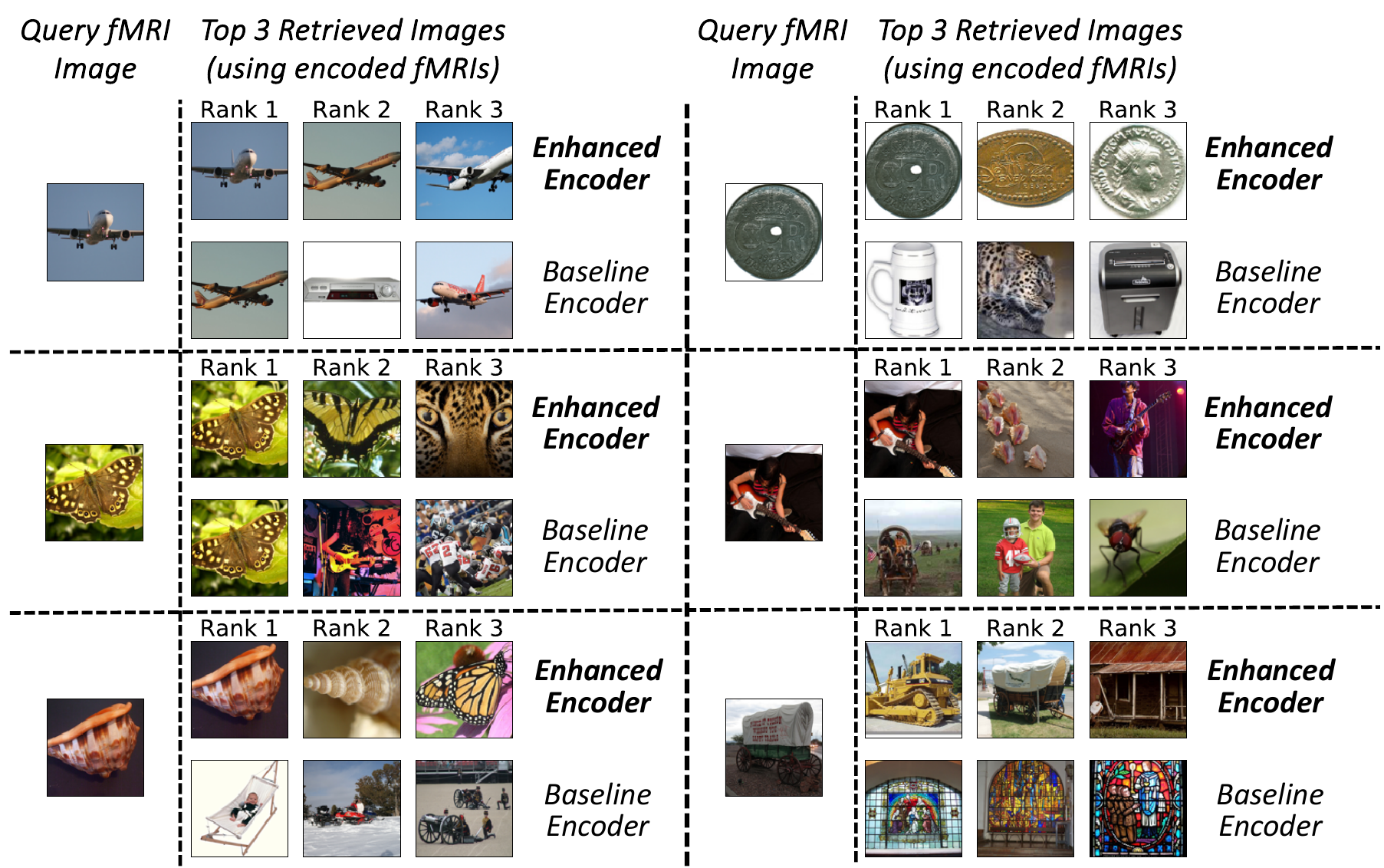}
     \end{subfigure}
     \begin{subfigure}{\textwidth}
         \centering
         \includegraphics[width=\textwidth]{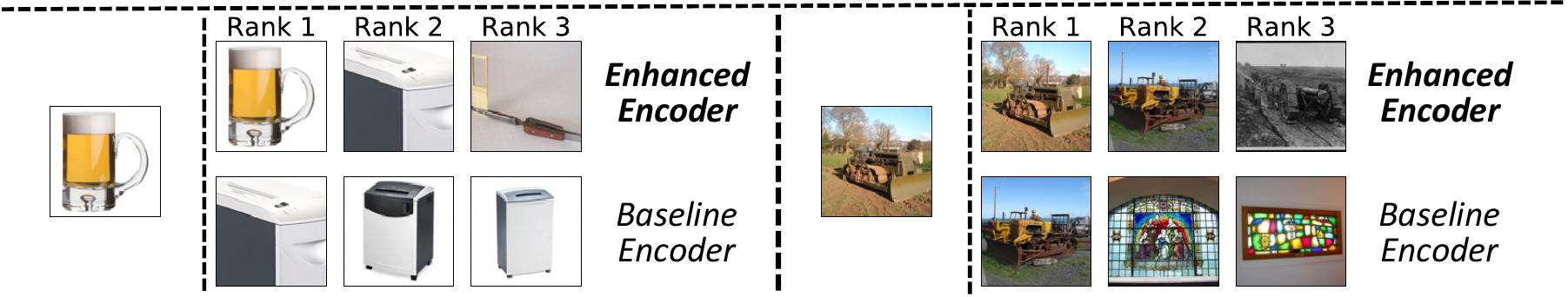}
     \end{subfigure}

 \caption{\textbf{Image Retrieval via encoded fMRI}:
 {\textbf{(a)} Image retrieval accuracy based on encoded fMRI, showing top-1 accuracy by comparing the test fMRI to 1000 encoded fMRIs, averaged across all subjects and for individual subjects. The image retrieval via fMRI predicted by a B2B-enhanced encoder (Orange line -- with a strong NSD teacher), is typically 200\%-500\% more accurate than retrieval from an fMRI predicted by the Baseline encoder (Grey line -- same encoder, but without a teacher). \textbf{(b)} The figure shows the top 3 retrieved images for each test fMRI, comparing the baseline encoder with the enhanced encoder trained using an NSD teacher. For each test case, 1000 candidate images were used, and the top 3 matches are presented.
 }
 }
  \label{Figure:Retrieval}

\end{figure*}

{In all the results presented next, }{\emph{grey color}} represents the student baseline encoder (trained solely on the student's fMRI data), \emph{purple color} illustrates the improvement using a teacher from the same dataset (GOD), and \emph{orange color} demonstrates the greater improvement achieved with a teacher from the 7-Tesla dataset (NSD). The ``number of training examples" refers to the student's own image-fMRI pairs, while the teacher encoder (trained on all available examples) remains unchanged during the student’s training process.

\vspace{0.15cm}
\noindent
{\underline{\textbf{(i)~Pearson Correlation Evaluation:}}} \vspace{2.5pt}
\cref{Figure:Encoder_GOD}.a shows {mean Pearson correlation results}
for individual subjects (Subject2 and Subject3), and averaged over all five subjects in GOD dataset (showing mean values and SEM {on GOD test set}). Those graphs show that Image-to-fMRI encoders trained with a ``teacher" subject {(via our B2B transformation),} exhibit a significant improvement over their baseline encoder {(trained only on the ``student"'s data)}. Notably, employing Subject4 from the same dataset as a ``teacher" results in a substantial improvement of approximately 50\% { over the baseline {encoder} (\(N=48\), \(p<0.00001\), \(d=4.703\)).} Remarkably, utilizing a subject from the higher-resolution NSD dataset as a ``teacher" leads to even more pronounced performance gains. Specifically, for a small number of examples, it demonstrates a nearly 100\% relative improvement over the baseline encoder, and {10\%-30\%} over the other enhanced encoder model {(\(N=48\), \(p<0.0001\), \(d=3.143\))}. These observations are remarkable given that the NSD dataset does not share any images with the GOD dataset  {and was scanned on a different fMRI machine}, whereas a ``teacher" subject from the same dataset shares numerous images with the ``student" subject {(and was scanned on the same fMRI machine)}. 
 Supplementary examples showcasing the improvement of all specific subjects' encoders are provided in~\cref{SM_Figure:Encoder_GOD} in the Supplementary Material.
 
 {\cref{Figure:Encoder_GOD}.b {further} presents results of enhancing Image-to-fMRI encoding} on the old Vim-1 fMRI dataset. This dataset contains only two subjects who viewed grey-scale images. Given the {small number of subjects}, we showcase the method using a ``teacher" subject from other datasets, namely GOD or NSD. The plots show that encoders trained with a ``teacher" subject {(whether from GOD or NSD)} exhibit an improvement over their baseline Vim-1 encoder model. {Moreover, as expected, a ``teacher" from the NSD 7-Tesla dataset outperforms a ``teacher"} from the GOD 3-Tesla dataset.
Notably, despite the considerable differences between the ``teacher" and ``student" subjects in terms of data characteristics (Color vs Grey-scale images, {and different fMRI machines}), the method remains effective in leveraging one subject's higher-quality data to improve another.

\vspace*{0.2cm}
\noindent
{\underline{\textbf{(ii)~Image Retrieval via Encoded fMRI:}}} \vspace{2.5pt}
{We further assess the impact of our B2B-enhanced encoding through an fMRI retrieval test. For each fMRI scan in the test set {(denoted as ``Query'')}, we aim to retrieve (detect) its corresponding \emph{Test-image} which produced it out of a set of N images (the Test-image and  N-1 random distractors).
To do so, we first predict the fMRIs of all N images in the set (using either the Baseline encoder or the B2B-enhanced encoder).
We then search for the Nearest-Neighbor (NN) of the \emph{real} test-fMRI (Query) among the set of N \emph{predicted} fMRIs  (using cosine similarity). If the fMRI predicted from the Test-image is retrieved as the $1^{st}$ NN, it obtains a ``Rank-1'' score. If it is retrieved as the $k^{th}$ NN, it obtains a ``Rank-k'' score.  The reported Top-1 accuracy is the percent of test-fMRIs which obtained a Rank-1 score.}

{ \cref{Figure:Retrieval}.a shows significant improvement in retrieval accuracy of GOD ``student'' encoders, when enhanced with a ``teacher" either from the GOD dataset or from the NSD dataset. Each Test-fMRI is compared against 1000 \emph{encoded} fMRIs -- from the real Test-image, and from 999 other `distractor' images from ImageNet. The B2B-enhanced encoder, especially when using an NSD teacher, outperforms the Baseline encoder.} 
{Please note that image retrieval via fMRI predicted by a \emph{B2B-enhanced encoder}  (Orange line -- with a strong NSD teacher), is typically \emph{200\%-500\% more accurate} than retrieval from an fMRI predicted by the \emph{Baseline encoder} (Grey line -- \emph{same encoder}, but without a teacher). This shows the significant power of our B2B transformations across brains and datasets, despite having no shared data.}

{\cref{Figure:Retrieval}.b visually presents qualitative retrieval results comparing the Baseline Encoder to the B2B-Enhanced Encoder (trained with the NSD ``teacher''). For each test fMRI, we show the top 3 retrieved images (rank 1, 2, and 3) out of 1000.  As seen, the B2B-Enhanced encoder not only retrieves the correct image more accurately (lower rank), but also the $2^{nd}$ and $3^{rd}$ ranked images are often from the same semantic category or visually similar to the correct image. This suggests that the B2B-Enhanced encoder generates fMRIs which capture more nuanced similarities, keeping related images closer in fMRI space. This further demonstrates its effectiveness in better representing brain responses.}

\vspace*{0.2cm}
\noindent
{\underline{\textbf{(iii)~Image  Classification from fMRI:}}} \vspace{2.5pt}
{We further evaluated the significance of our enhanced fMRI encoding for ``Zero-Shot" classification -- namely, classification of fMRI data into \emph{new image categories} that were never seen during training. 
{We devised a simple fMRI-classification algorithm, which uses an Image-to-fMRI encoder to \emph{predict} fMRI patterns for 100 arbitrary images per class (since these are never-before seen image categories for which there are no real fMRI \emph{training} examples). These 100 predicted fMRIs are then} averaged to form an ``fMRI Representative'' per category. In the classification stage, each test-fMRI (a real fMRI scan) is compared against these generated fMRI class representatives using cosine similarity. The most similar fMRI-representative determines the category classification.  In an $n$-way} classification task, the test-fMRI  is compared against $n$ fMRI class-representatives: the \emph{generated} fMRI-representative of the \emph{correct} image class, along with  $n$-1 generated fMRI-representatives of other $n$-1 randomly selected ImageNet classes.

\begin{figure*}
     \centering
     \begin{subfigure}{\textwidth}
         \centering
         \includegraphics[width=\textwidth]{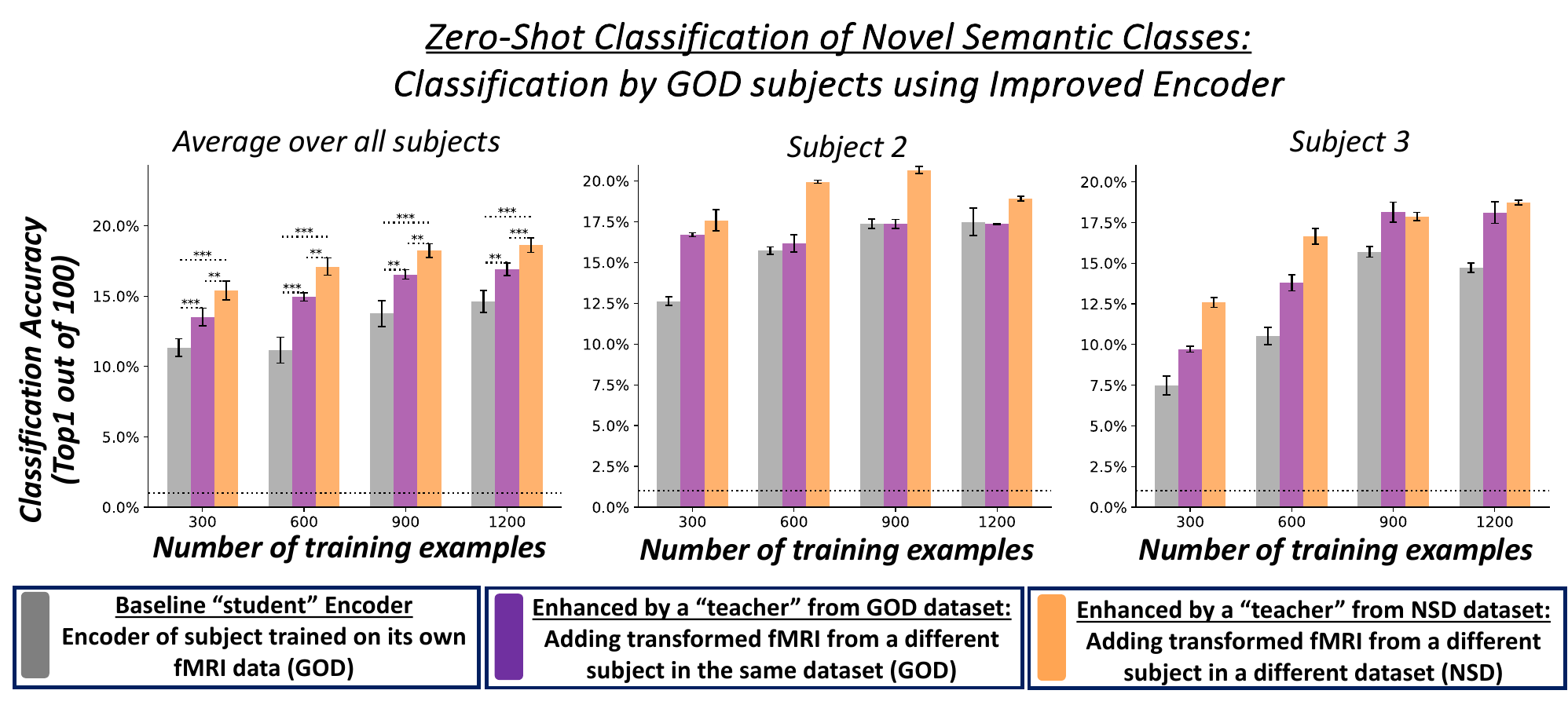}
     \end{subfigure}
\caption{\textbf{Image Classification from fMRI of 3-T GOD subjects:}
This plot presents the classification {accuracy (out of 100 classes) using Baseline vs. B2B-enhanced encoder}. The results presented are the mean accuracy of all subjects with the SEM error {and individual subject result with mean and SEM over 3 training repetitions. The image classification from fMRI predicted by a B2B-enhanced encoder (Orange bar -- with a strong NSD teacher), is typically $\sim$20\%-50\% more accurate than image classification from an fMRI predicted by the Baseline encoder (Grey bar -- same encoder, but without a teacher), and 1500\%-2000\% more accurate than chance level (Grey dashed-line). The $n$=100 image classes/categories were never seen during training time, neither by the teacher, nor by the baseline encoder.}.}
\label{Figure:Classification}
\end{figure*}

\cref{Figure:Classification} compares the $100$-way classification  {accuracy (out of $n$=100 classes)} using {Baseline vs. B2B-enhanced} encoders. The results indicate that, irrespective of the number of training examples, the B2B-enhanced encoders (represented by the purple column) achieve significantly higher classification accuracy {(paired samples permutation test, \(N=20\), \(p<0.00001\), \(d=1.01\))} than the baseline encoders (grey column). Furthermore, the B2B-enhanced encoder trained with the 7-Tesla {``teacher"} (orange column) demonstrates even greater classification accuracy than the GOD 3-Tesla ``teacher" {(\(N=20\), \(p<0.00001\), \(d=0.98\)).}
{In particular, note that image classification from fMRI predicted by a \emph{B2B-enhanced encoder}  (Orange bar -- with a strong NSD teacher), is typically \emph{\textbf{$\sim$20\%-50\% more accurate}} than image classification from an fMRI predicted by the \emph{Baseline encoder} (Grey bar -- same encoder, but without a teacher), and \emph{\textbf{1500\%-2000\% more accurate}} than \emph{chance level} (Grey dashed-line). Recall that all $n$=100 image classes/categories were never seen during training time, neither by the teacher, nor by the baseline encoder.}
\\

\vspace{-0.2cm}
\noindent
{\emph{\textbf{Analysis and ablations}} of the various factors which contribute to the improved encoder performance are provided in the Supplementary \cref{sec:ablation}.} {In particular, it analyzes the influence of data quality and quantity, showing that both factors contribute to NSD's superiority as a dataset for enhancing other subjects' encoders.
}

\newpage
\section{Discussion}

This paper introduces, for the first time, an approach to learn functional brain-to-brain {(B2B)} transformations between different subjects without requiring any shared data.  {Our approach  enables the merging and enrichment of diverse fMRI datasets {collected around the globe}, even when they were collected on different image stimuli and on different fMRI machines of varying resolutions.} {This may provide a significant boost in brain research, by increasing both the quantity and quality of fMRI data available for analysis.}
We {show that it is possible to} utilize both ``non-shared" and ``external" images to train {functional B2B transformations, and to significantly improve Image-to-fMRI encoders.} This ability to incorporate images unseen by any of the subjects marks a significant advancement, greatly expanding the 
{number of available training data (by several orders of magnitude).}
Our method is particularly effective in utilizing new high-quality fMRI datasets, such as the NSD 7-Tesla dataset, to enhance {Image-to-fMRI encoders of subjects in} older, lower-quality datasets like the {GOD (3-Tesla) and Vim1 (4-Tesla)} datasets. We have demonstrated {that these \emph{B2B-Enhanced Encoders} further improve the accuracy of ``zero-shot'' semantic image classification  directly from fMRI, 
as well as provide improved fMRI retrieval accuracy.}

{Our method {was so far} demonstrated effectively {on} fMRI data in response to visual stimuli{.} Extending our approach to {to other \emph{sensing modalities} (e.g., MEG, EEG, ECoG), or to other types of \emph{data stimuli} (e.g., audio, video, text) has great potential.} {As with any Deep-Learning based method, the performance of our approach may deteriorate} when applied to out-of-distribution images that differ significantly from the training set. {Therefore, both our B2B transformation and enhanced encoder may have limited applicability when datasets differ significantly in image characteristics.} Despite this, there are likely underlying similarities in image features and corresponding brain activations that our method can leverage, although with potentially reduced effectiveness. 
Finally, the quality of the encoder is a key factor in our method’s success. While small datasets with limited number of samples per subject can limit the overall quality of the results, our findings indicate that our method can still have good prediction correlation with as few as 300 image-fMRI example pairs.}

{Our capability to aggregate data across diverse datasets and conditions without requiring shared data, alongside improvements in encoder performance, provides unique opportunities for advancing cognitive neuroscience and computational neuroimaging. By facilitating direct comparisons between subjects who viewed different image sets, recorded on various machines in labs worldwide, our B2B-transformation approach opens the door to broader, cross-dataset analyses which were previously unattainable, with many new potential applications. For example, our B2B-transformations may facilitate the integration of fMRI datasets of different \emph{behavioral tasks} across different fMRI datasets. It may create new opportunities to study a range of visual and cognitive tasks \emph{collectively} — a level of analysis challenging, if not impossible, without a mechanism to bridge data collected under varied conditions. Our approach may also support future studies of brain functionality across individuals with diverse backgrounds or health conditions, whether they are viewing images or, potentially, engaging with other stimuli types like video and audio.}

{Our B2B-enhanced Image-to-fMRI encoders  also open the door to many new potential opportunities. Their effective performance with minimal data may be particularly valuable for labs that cannot collect extensive datasets but aim to investigate specific scenarios. Due to the encoder’s ability to predict fMRI activations on a vast number of images, researchers may explore and compare fMRI responses to a wider variety of stimuli. For example, compare fMRI responses to images of indoor versus outdoor scenes, or even contrast fMRI responses to emotional content (e.g., happy versus sad images).   
All in all, our methods for training B2B transformations without shared data and using those to enhance Image-to-fMRI encoding could serve as valuable tools for future research.}

\section{Methods}

\subsection{Datasets}
\label{Methods:Datasets}
We tested our approach on three distinct publicly available  fMRI datasets: ``Natural Scenes Data set`` (NSD)~\cite{allen2022massive}, Generic Object Decoding (GOD)~\cite{Horikawa2017GenericFeatures}, and ``Visual Imaging 1" (Vim1)~\cite{kay2008identifying,naselaris2009bayesian}. 
The datasets provide fMRI responses of human subjects to a variety of natural images.  We summarize differences between the datasets in the table below.

 Natural Scenes Dataset (NSD) is a new dataset recorded using 7-Tesla fMRI machine, resulting in higher voxel resolution. All 8 subjects in this dataset were presented with 1,000 shared images and 9,000 unique images per subject. This is contrary to other datasets where all participants were presented the same stimuli. Vim1 dataset has grey scale stimuli images with a circular mask (see Figure \ref{3}), whereas the other datasets that have RGB images. It should be noted that each dataset was scanned with a different machine as well as different recording scheme and processing protocols. We use the same approach and architecture to produce results for all the datasets. 
 In addition to the fMRI datasets, we also used 50,000 natural images as ``External data" (images without any fMRI). Those images were taken from 1000 classes of ImageNet validation data (``ILSVRC"~\cite{Deng2009ImageNet:Database}). Those images are the ``external" data reported in our experiments.
{For both the VIM-1 and GOD datasets, we utilized the predefined test sets, which consist of 120 images for VIM-1 and 50 images for GOD. We reserved 10\% of the training set to serve as a validation set. In the case of NSD, since there was no dedicated test set, we allocated 10\% of the data for validation and another 10\% for testing. Both validation and test sets were proportionally sampled from the shared data across all subjects and the non-shared data. This approach was crucial to ensure that the test set included shared data, enabling the evaluation of transformations that can only be assessed using shared data across subjects.}
 
\begin{table*}[ht!]
\centering
\begin{tabular}{ c|c|c|c|c|c|c|c } 
\textbf{Dataset} & \textbf{Number of distinct stimuli} & \textbf{Subjects}  & \textbf{Voxels}  & \textbf{Resolution} & \textbf{Image origin}
\\ 
\hline
NSD~\cite{allen2022massive} & 10000 per subject ($\sim{73000}$\text{ total}) & 8  & 10000 & 7T & COCO ~\cite{cocodataset}
\\
GOD~\cite{Horikawa2017GenericFeatures} & 1250 per subject (1250 total) & 5  & 5000 & 3T  & ImageNet ~\cite{Deng2009ImageNet:Database}
\\ 
Vim1~\cite{kay2008identifying,naselaris2009bayesian} & 2000 per subject (2000 total) & 2 & 5000 & 4T & Multiple sources

\end{tabular} \\ [0.5ex]
\label{table:DatasetsSummary}
\end{table*}

\subsection{Data Acquisition and Processing}
\label{Sec:methods_data}
{The datasets utilized in our study include BOLD fMRI responses to various natural images, recorded over multiple scanning sessions. In all experiments, participants were instructed to fixate their eyes on the center of the presented images. Each dataset underwent specific pre-processing procedures as detailed in their respective publications~\cite{kay2008identifying,naselaris2009bayesian,Horikawa2017GenericFeatures,allen2022massive}. Building on these procedures, we implemented additional processing steps:}

\paragraph{Voxel Selection.} {Across all datasets, a subset of brain voxels relevant to visual stimuli was selected. For the GOD dataset, we utilized a predefined selection of 5,000 voxels from the visual areas. In the NSD and VIM-1 datasets, we applied a data-driven approach to voxel selection based on the signal-to-noise ratio (SNR). SNR was computed as the ratio of the variance in voxel responses to different stimuli (reflecting true neural activity) to the variance in responses to the same stimulus across multiple repetitions (reflecting noise). In NSD, each image had between one and three repetitions, and in VIM-1, each image had two repetitions in the training set.
By selecting the top voxels with the highest SNR, we ensured that the chosen voxels were those that most reliably reflected neural responses to visual stimuli. 
Specifically, we selected the top 5,000 voxels for VIM-1 to align with the GOD dataset, and the top 10,000 voxels for NSD , reflecting its higher resolution and the greater number of scanned images. In both dataset voxels were chosen from the entire brain.}

\paragraph{Voxel Z-scoring.} Each voxel underwent Z-scoring normalization within each run, defined as a continuous fMRI scanning period. This normalization standardized voxel responses across runs, improving data comparability and consistency.

\paragraph{Averaging across repetitions.} {To enhance fMRI data quality, images were presented multiple times to each subject, allowing for higher-quality measurements. In VIM-1, each training image was repeated twice, and each test image 13 times. NSD images had between one and three repetitions. We averaged the fMRI responses across repetitions to improve signal quality. For the GOD dataset, we used a version with five repetitions per image in the training set and 24 in the test set. To illustrate the impact of repetition, we present in the supplementary results using both single-repetition fMRI and averaged fMRI with varying repetition counts.}

\paragraph{Group space fMRI - fsaverage.} For comparative analysis between anatomical and functional alignments, we used the 'fsaverage' shared space provided by FreeSurfer~\cite{fischl2012freesurfer,dale1999cortical} only for the transformation evaluation in the NSD dataset. This approach enabled a fair and direct comparison by performing both alignments in a common reference space. Specifically, we selected a consistent set of 10,000 voxels across all subjects, based on the SNR of Subject 1. The fsaverage space was derived using pre-calculated betas from the NSD This procedure involved resampling subject-native cortical surfaces at three different depths using cubic interpolation. Subsequently, the averaged betas were produced by averaging these resampled values. Finally, these averaged betas were mapped onto the fsaverage surface through nearest-neighbor interpolation. This methodical approach allows us to compare functional alignment results directly with those obtained from anatomical alignment, each within the same standardized brain space.

\subsection{Visual Encoder}

The image-to-fMRI encoder employed in our work is based on the framework outlined in~\cite{gaziv2022self}. It encompasses a pre-trained VGG network (trained on ImageNet classification task), augmented with additional convolution layers and a non-linear layer to project data into the fMRI space. First, the input image is passed through VGG19 network~\cite{simonyan2014very}, intermediate embeddings of the network blocks 1-4 are extracted, each compromising a parallel branch of feature representation. Each of the branches feature map is reduced spatially and the channels' dimension as well to 28$\times$28$\times$32 (Height$\times$Width$\times$ConvolutionChannels). This is done by a series of modules consisting of 3$\times$3 convolution with 32 channels, ReLU, $\times$2 sub-sampling, and batch normalization. 
On the reduced feature maps, we train a spatial layer that transfers the spatial position of the feature map to voxels. For each voxel, we learn a linear combination of the spatial positions (where channels are not affected). This is done for each one of the branches and the layer weights are shared between the branches. After this operation we have a representation with dimensions Voxels$\times$ConvolutionChannels.
This is followed by a locally connected layer with output size of 1, this layer learns how to combine the channels for each one of the voxels, representation with dimensions Voxels$\times$1, for each one of the branches.
 The final layer combines the results of the 4 branches, this is done by concatenating the 4 branches and applying a locally connected layer with positive weights (averaging).

\subsection{Training brain-to-brain transformations with no shared data}

The brain-to-brain transformation at the core of our method is implemented via a simple neural architecture featuring a single linear layer {mapping all voxels of one subject onto each voxel of the other subject}, optimized with an L2 regularization loss (detailed below).  Prior to training the transformation network $T$, we performed separate training for the visual encoder of each subject individually (on their individual data), using the procedure recommended in~\cite{gaziv2022self} (refer to Figure~\ref{2}.a). Notably, during transformation training, the encoder weights remained fixed.

The transformation $T$ serves the purpose of mapping fMRI patterns from one subject (Subject1) to their corresponding patterns in another subject (Subject2).  Leveraging the predictive capability of the pre-trained encoders for forecasting fMRI responses to images, we are capable of generating corresponding fMRI patterns for ``non-shared" and ``external" images. These are used to train the transformation $T$, without any shared data between the two subjects. Our fMRI loss function adopts the mean square error loss between fMRIs. {The loss is calculated between the target measured fMRI of Subject2 and the predicted fMRI after applying the transformation $T$ from Subject1 to Subject2}.

\paragraph{Losses.} {} 
Our overall training loss encompasses three components corresponding to different data types (shared, non-shared, and external), and in each scenario, we can use all or only some of them together. Each component employs the same fMRI loss metric, with variations in the types of fMRI data compared (shared, non-shared, or external) and the coefficients assigned. A regularization term $\mathcal{L}_{Reg}$ employs L2 regularization on the weights of the linear transformation layer, aligning with prior methods~\cite{yamada2015inter}. The total loss, can be represented as follows:

\[ \mathcal{L} = \alpha_S \, \mathcal{L}_S + \alpha_{NS} \, \mathcal{L}_{NS} + \alpha_{Ext} \, \mathcal{L}_{Ext} + \alpha_{Reg} \, \mathcal{L}_{Reg} \]

The first loss term, $\mathcal{L}_S$, leverages ``shared data":
{\[ \mathcal{L}_S =  \mathcal{L}_{fMRI}({T}_{1,2}({\mathcal{F}}_{1}(\mathcal{I}_{Shared})),\mathcal{F}_{2}(\mathcal{I}_{Shared})) \]}

where $\mathcal{F}_{1}$ represents the recorded fMRI of Subject1, and $\mathcal{F}_{2}$ represents the recorded fMRI of Subject2, while observing the same image {$\mathcal{I}_{Shared}$}. ${T}_{1,2}$ denotes the transformation model mapping from Subject1 to Subject2. We employ the fMRI loss of the transformation to compare the transformed fMRI with the actual measured fMRI.

The second loss term, $\mathcal{L}_{NS}$, utilizes ``non-shared data" (refer to Figure~\ref{2}.b):
{\[ \mathcal{L}_{NS} =  \mathcal{L}_{fMRI}(T_{1,2}(\mathcal{F}_{1}(\mathcal{I}_{N_{s_1}})),E_2(\mathcal{I}_{N_{s_1}})) \ \ or \ \ \mathcal{L}_{NS} = \mathcal{L}_{fMRI}(T_{1,2}(E_1(\mathcal{I}_{N_{s_2}})),\mathcal{F}_{2}(\mathcal{I}_{N_{s_2}})) \]}

where {$\mathcal{F}_{1}(\mathcal{I}_{N_{s_1}})$ and $\mathcal{I}_{N_{s_1}}$ are the measured} fMRI of Subject1 and the corresponding image (the image viewed exclusively by Subject1). We transform Subject1's fMRI signals and compare them to the encoded fMRI signals derived from the image seen by Subject1 (using Subject2's encoder). In the same manners, {$\mathcal{F}_{2}(\mathcal{I}_{N_{s_2}})$ and $\mathcal{I}_{N_{s_2}}$} correspond to the recorded fMRI of Subject2 and the corresponding image (the image viewed exclusively by Subject2). We transform the encoded fMRI signals derived from the image seen by Subject2 (using Subject1's encoder) and compare them to Subject2's fMRI signals $\mathcal{F}_2$.

The third loss term, $\mathcal{L}_{Ext}$, involves ``External" images (refer to Figure~\ref{2}.c):
\[ \mathcal{L}_{Ext} =  \mathcal{L}_{fMRI}({T}_{1,2}(E_1({I}_{Ext})),E_2({I}_{Ext}))\]
\noindent
In this case, ${I}_{Ext}$ represents an external image with no corresponding fMRI data for any of the subjects. We feed this image into the encoders of both subjects. After encoding the fMRI, the loss is similar to that in the case of $\mathcal{L}_{S}$, utilizing the encoded fMRI.

It is worth noting that the number of shared and non-shared fMRI data is constrained by the available data, while the number of external data is essentially limitless. In occasions where we train the transformation without any shared data, either within or between datasets, we omit the first loss term $\mathcal{L}_S$ and employ only the other two loss components ($\mathcal{L}_{NS}$ and $\mathcal{L}_{Ext}$).

\paragraph{Training Technical Details.} {}
{The training is conducted using scikit-learn's Ridge regression model using the different kind of data types. The samples are weighted according to the loss coefficients mentioned above ($\alpha_S$, $\alpha_{NS}$ and $\alpha_{Ext}$). We separate the fMRI data into 3 sets: Train (80\%), Validation (10\%) and Test (10\%). Subsequently, we select the optimal regularization term by assessing the transformation model performance on the validation set. When training with non-shared fMRI data we also used 6000 external images. We used $\alpha_S=1, \alpha_{NS}=0.5$ and $\alpha_{Ext}=0.1$}.

\subsection{Improving Image-to-fMRI Encoding via B2B Transformation}

We enhance the image-to-fMRI encoder of one subject using another subject with superior data quality or more extensive examples. For simplicity, we term this technique the ``teacher-student" method, where the subject with higher quality data plays the role of the ``teacher", and the poorer-quality subject is the ``student".

The process initiates by training the ``teacher" subject's encoder ($E_{t}$) using its complete dataset, with the resulting weights remaining unchanged. Following this, we engage in simultaneous training of the ``student" subject's encoder ($E_{s}$) and the transformation responsible for mapping fMRI data between the ``teacher" and ``student" subjects ($T_{t,s}$). This joint training benefits both the encoder and the transformation, allowing them to improve collaboratively. Importantly, during this step, the ``student" encoder is not solely trained on its own data but also incorporates information from the ``teacher" encoder and its unique (often non-shared) fMRI data, as well as ``external" data (images without any fMRI).

\paragraph{Losses.} {} Our comprehensive loss function comprises three components corresponding to different data types (shared, non-shared, and external) {used for training both the student encoder and the transformation}. In each scenario, we can choose to use all or only some of these components together. Additionally, we include an encoding loss, which pertains to the student encoder and its available data. Lastly, a regularization term $\mathcal{L}_{Reg}$ employs L2 regularization on the weights of the linear transformation layer.  Our loss function can be represented as follows:
\[ \mathcal{L} = \alpha_{Enc} \, \mathcal{L}_{Enc} +\alpha_{S} \, \mathcal{L}_{S} + \alpha_{NS} \, \mathcal{L}_{NS} + \alpha_{Ext} \, \mathcal{L}_{Ext} + \alpha_{Reg} \, \mathcal{L}_{Reg} \]

The first loss, $\mathcal{L}_{Enc}$, calculates the student encoding loss using all the available image-fMRI pairs of the student subject:
\[ \mathcal{L}_{Enc} =  \mathcal{L}_{fMRI}(E_{Student}({I}_{Student}),{F}_{Student}({I}_{Student}))\]
\noindent
In this equation, ${I}_{Student}$ and ${F}_{Student}$ represent the image-fMRI pairs of the student subject. The loss measures the discrepancy between the encoded image (by the student encoder) and the actual measured fMRI using the fMRI encoding loss.

The second loss, $\mathcal{L}_{S}$, is similar to the shared data loss from the previous section. It calculates the transformation loss based on all the shared fMRI-fMRI pairs in the dataset, corresponding to both the student and the teacher.

Subsequently, we have two additional losses, $\mathcal{L}_{NS}$ and $\mathcal{L}_{Ext}$ (see Figure~\ref{2}.b and~\ref{2}.c), calculated in a manner similar to that presented in the previous section, with the distinction that the student encoder weights are not fixed and are affected by the training process. For instance, if we consider the scenario of $\mathcal{L}_{NS}$, we take the actual fMRI response $F_{Teacher}$ from the ``teacher" corresponding to an image $I_{Teacher}$ that the ``student" has never seen before. This response is then transformed to the ``student's" fMRI space and compared to the encoded response produced by the ``student" encoder for the same image. Namely: 
\[ \mathcal{L}_{NS} =  \mathcal{L}_{fMRI}(E_{Student}({I}_{Teacher}),T_{T,S}({F}_{Teacher}({I}_{Teacher})))\]
Similarly for any external image $I$:
\[ \mathcal{L}_{Ext} =  \mathcal{L}_{fMRI}(E_{Student}({I}_{Ext}),{T}_{T,S}(E_{Teacher}({I}_{Ext})))\]

\paragraph{Training Technical Details.} {}
{ The training is conducted using the Adam optimizer for  30 epochs with an initial learning rate of 5e-4 (employing step decay) to update the weights of both the encoder and the transformation model. Each training batch consists of 32 examples from each type: 32 image-fMRI pairs from the student subject, 32 image-fMRI pairs from the teacher subject for non-shared data, 32 pairs for shared data (if available) and another 32 external images (which are encoded using both the teacher and student encoders). In each batch, different pairs from the teacher and different external images are sampled to ensure variety and comprehensive training.} To determine the optimal coefficients for the loss functions and regularization, we employ a validation procedure, as outlined in a previous section. When applying our method to the ``Vim1" dataset and the NSD dataset, we encountered a few challenges. These datasets differ in terms of color representation, with one being grey-scale and the other in RGB format. To bridge this domain gap, we converted the external and non-shared images to grey-scale, minimizing disparities between them.

{The selection of the teacher subject is a critical step in our method. We choose the teacher based on its effectiveness in enhancing the encoding performance of other subjects on a validation set. Importantly, the teacher can be chosen from within the same dataset or from a different dataset, depending on the specific requirements of the application.}

\subsection{Decoding model}
\label{sec:method_decoding}
{In this study, we employed a decoding process to reconstruct images from fMRI data following the completion of the transformation. Our decoding process combines two distinct methods, which are detailed in~\cite{gaziv2022self} and~\cite{scotti2023reconstructing}.
First, we utilized the model introduced in~\cite{gaziv2022self}. This model, when provided with an fMRI input, generates a reconstructed image that closely aligns with the structural aspects of the original image. However, the output may not appear entirely natural. To address this limitation, we incorporated a novel decoding model from~\cite{scotti2023reconstructing}. This model leverages a diffusion-based approach to produce highly realistic natural images from fMRI data. Nevertheless, it may not preserve the fine structural details as effectively.
To strike a balance between fidelity to the original image and realism, we adopted a two-step decoding strategy. Initially, we employed the first model to generate a reconstructed image, which faithfully captures the structural aspects of the original but might lack naturalness. Then, we used this initial reconstruction as the starting point for the second model, which refines the image to make it both faithful to the original and visually realistic. This combined approach allows us to achieve a compelling balance between structural accuracy and natural appearance in the reconstructed images.}

\subsection{Statistical tests}
{The statistical test that was used throughout this work is a two-way paired permutation test with a one-sided alternative. The number of permutations that was used is 100000, and the statistic calculated is the mean of the differences. This was done using the stats.permutation\_test function in the python package scipy. In figures, * marks \(p<0.05\), ** marks \(p<0.01\) and *** marks \(p<0.001\), and the dotted lines mark which models were compared.}

\section{Data availability}
All the datasets analysed during the current study are publicly available. The NSD dataset is available through \url{https://naturalscenesdataset.org} with access agreement submission. The GOD dataset is available in the ``GenericObjectDecoding" repository, \url{https://github.com/KamitaniLab/GenericObjectDecoding}. The Vim1 dataset is available in \url{https://crcns.org/data-sets/vc/vim-1/about-vim-1}.

\section{{Code availability}}
{The code for both the brain-to-brain transformation and the enhanced encoder methods, along with scripts used to create the figures for the paper, is available on GitHub at \\ \url{https://github.com/navvewas/Brain-Transformation-with-No-Shared-Data.git}}

\section{Declaration of generative AI and AI-assisted technologies in the writing process}
During the preparation of this work the authors used ChatGPT in order to improve the readability and language of the manuscript. After using this tool/service, the authors reviewed and edited the content as needed and take(s) full responsibility for the content of the published article.

\section{Acknowledgments}
This research was supported by the European Research Council (ERC) under the Horizon program, grant number 101142115.

\bibliography{main}

\begin{thebibliography}{10}

\bibitem{kanwisher1997fusiform}
Nancy Kanwisher, Josh McDermott, and Marvin~M Chun.
\newblock The fusiform face area: a module in human extrastriate cortex specialized for face perception.
\newblock {\em Journal of neuroscience}, 17(11):4302--4311, 1997.

\bibitem{epstein1998cortical}
Russell Epstein and Nancy Kanwisher.
\newblock A cortical representation of the local visual environment.
\newblock {\em Nature}, 392(6676):598--601, 1998.

\bibitem{downing2001cortical}
Paul~E Downing, Yuhong Jiang, Miles Shuman, and Nancy Kanwisher.
\newblock A cortical area selective for visual processing of the human body.
\newblock {\em Science}, 293(5539):2470--2473, 2001.

\bibitem{tang2017using}
I-Chun Tang, Yu-Ping Tsai, Ying-Ju Lin, Jyh-Horng Chen, Chao-Hsien Hsieh, Shih-Han Hung, William~C Sullivan, Hsing-Fen Tang, and Chun-Yen Chang.
\newblock Using functional magnetic resonance imaging (fmri) to analyze brain region activity when viewing landscapes.
\newblock {\em Landscape and Urban Planning}, 162:137--144, 2017.

\bibitem{heeger2002does}
David~J Heeger and David Ress.
\newblock What does fmri tell us about neuronal activity?
\newblock {\em Nature reviews neuroscience}, 3(2):142--151, 2002.

\bibitem{riddle1995individual}
DR~Riddle and Dale Purves.
\newblock Individual variation and lateral asymmetry of the rat primary somatosensory cortex.
\newblock {\em Journal of Neuroscience}, 15(6):4184--4195, 1995.

\bibitem{frost2012measuring}
Martin~A Frost and Rainer Goebel.
\newblock Measuring structural--functional correspondence: spatial variability of specialised brain regions after macro-anatomical alignment.
\newblock {\em Neuroimage}, 59(2):1369--1381, 2012.

\bibitem{conroy2013inter}
Bryan~R Conroy, Benjamin~D Singer, J~Swaroop Guntupalli, Peter~J Ramadge, and James~V Haxby.
\newblock Inter-subject alignment of human cortical anatomy using functional connectivity.
\newblock {\em NeuroImage}, 81:400--411, 2013.

\bibitem{zhen2015quantifying}
Zonglei Zhen, Zetian Yang, Lijie Huang, Xiang-zhen Kong, Xu~Wang, Xiaobin Dang, Yangyue Huang, Yiying Song, and Jia Liu.
\newblock Quantifying interindividual variability and asymmetry of face-selective regions: a probabilistic functional atlas.
\newblock {\em Neuroimage}, 113:13--25, 2015.

\bibitem{mazziotta2001probabilistic}
John Mazziotta, Arthur Toga, Alan Evans, Peter Fox, Jack Lancaster, Karl Zilles, Roger Woods, Tomas Paus, Gregory Simpson, Bruce Pike, et~al.
\newblock A probabilistic atlas and reference system for the human brain: International consortium for brain mapping (icbm).
\newblock {\em Philosophical Transactions of the Royal Society of London. Series B: Biological Sciences}, 356(1412):1293--1322, 2001.

\bibitem{talairach19883}
J~Talairach.
\newblock 3-dimensional proportional system; an approach to cerebral imaging. co-planar stereotaxic atlas of the human brain.
\newblock {\em Thieme}, pages 1--122, 1988.

\bibitem{fischl2012freesurfer}
Bruce Fischl.
\newblock Freesurfer.
\newblock {\em Neuroimage}, 62(2):774--781, 2012.

\bibitem{dale1999cortical}
Anders~M Dale, Bruce Fischl, and Martin~I Sereno.
\newblock Cortical surface-based analysis: I. segmentation and surface reconstruction.
\newblock {\em Neuroimage}, 9(2):179--194, 1999.

\bibitem{haxby2011common}
James~V Haxby, J~Swaroop Guntupalli, Andrew~C Connolly, Yaroslav~O Halchenko, Bryan~R Conroy, M~Ida Gobbini, Michael Hanke, and Peter~J Ramadge.
\newblock A common, high-dimensional model of the representational space in human ventral temporal cortex.
\newblock {\em Neuron}, 72(2):404--416, 2011.

\bibitem{yamada2015inter}
Kentaro Yamada, Yoichi Miyawaki, and Yukiyasu Kamitani.
\newblock Inter-subject neural code converter for visual image representation.
\newblock {\em NeuroImage}, 113:289--297, 2015.

\bibitem{brett2002problem}
Matthew Brett, Ingrid~S Johnsrude, and Adrian~M Owen.
\newblock The problem of functional localization in the human brain.
\newblock {\em Nature reviews neuroscience}, 3(3):243--249, 2002.

\bibitem{guntupalli2016model}
J~Swaroop Guntupalli, Michael Hanke, Yaroslav~O Halchenko, Andrew~C Connolly, Peter~J Ramadge, and James~V Haxby.
\newblock A model of representational spaces in human cortex.
\newblock {\em Cerebral cortex}, 26(6):2919--2934, 2016.

\bibitem{chen2015reduced}
Po-Hsuan~Cameron Chen, Janice Chen, Yaara Yeshurun, Uri Hasson, James Haxby, and Peter~J Ramadge.
\newblock A reduced-dimension fmri shared response model.
\newblock {\em Advances in neural information processing systems}, 28, 2015.

\bibitem{de2010against}
Hans P~Op de~Beeck.
\newblock Against hyperacuity in brain reading: spatial smoothing does not hurt multivariate fmri analyses?
\newblock {\em Neuroimage}, 49(3):1943--1948, 2010.

\bibitem{thual2022aligning}
Alexis Thual, Quang~Huy Tran, Tatiana Zemskova, Nicolas Courty, R{\'e}mi Flamary, Stanislas Dehaene, and Bertrand Thirion.
\newblock Aligning individual brains with fused unbalanced gromov wasserstein.
\newblock {\em Advances in neural information processing systems}, 35:21792--21804, 2022.

\bibitem{robinson2014msm}
Emma~C Robinson, Saad Jbabdi, Matthew~F Glasser, Jesper Andersson, Gregory~C Burgess, Michael~P Harms, Stephen~M Smith, David~C Van~Essen, and Mark Jenkinson.
\newblock Msm: a new flexible framework for multimodal surface matching.
\newblock {\em Neuroimage}, 100:414--426, 2014.

\bibitem{lorbert2012kernel}
Alexander Lorbert and Peter~J Ramadge.
\newblock Kernel hyperalignment.
\newblock {\em Advances in Neural Information Processing Systems}, 25, 2012.

\bibitem{xu2012regularized}
Hao Xu, Alexander Lorbert, Peter~J Ramadge, J~Swaroop Guntupalli, and James~V Haxby.
\newblock Regularized hyperalignment of multi-set fmri data.
\newblock In {\em 2012 IEEE statistical signal processing workshop (SSP)}, pages 229--232. IEEE, 2012.

\bibitem{haxby2020hyperalignment}
James~V Haxby, J~Swaroop Guntupalli, Samuel~A Nastase, and Ma~Feilong.
\newblock Hyperalignment: Modeling shared information encoded in idiosyncratic cortical topographies.
\newblock {\em elife}, 9:e56601, 2020.

\bibitem{Horikawa2017GenericFeatures}
Tomoyasu Horikawa and Yukiyasu Kamitani.
\newblock {Generic decoding of seen and imagined objects using hierarchical visual features}.
\newblock {\em Nature Communications}, 8(1):1--15, 5 2017.

\bibitem{allen2022massive}
Emily~J Allen, Ghislain St-Yves, Yihan Wu, Jesse~L Breedlove, Jacob~S Prince, Logan~T Dowdle, Matthias Nau, Brad Caron, Franco Pestilli, Ian Charest, et~al.
\newblock A massive 7t fmri dataset to bridge cognitive neuroscience and artificial intelligence.
\newblock {\em Nature neuroscience}, 25(1):116--126, 2022.

\bibitem{yamins2014performance}
Daniel~LK Yamins, Ha~Hong, Charles~F Cadieu, Ethan~A Solomon, Darren Seibert, and James~J DiCarlo.
\newblock Performance-optimized hierarchical models predict neural responses in higher visual cortex.
\newblock {\em Proceedings of the national academy of sciences}, 111(23):8619--8624, 2014.

\bibitem{eickenberg2017seeing}
Michael Eickenberg, Alexandre Gramfort, Ga{\"e}l Varoquaux, and Bertrand Thirion.
\newblock Seeing it all: Convolutional network layers map the function of the human visual system.
\newblock {\em NeuroImage}, 152:184--194, 2017.

\bibitem{wen2018neural}
Haiguang Wen, Junxing Shi, Yizhen Zhang, Kun-Han Lu, Jiayue Cao, and Zhongming Liu.
\newblock Neural encoding and decoding with deep learning for dynamic natural vision.
\newblock {\em Cerebral cortex}, 28(12):4136--4160, 2018.

\bibitem{wen2018deep}
Haiguang Wen, Junxing Shi, Wei Chen, and Zhongming Liu.
\newblock Deep residual network predicts cortical representation and organization of visual features for rapid categorization.
\newblock {\em Scientific reports}, 8(1):3752, 2018.

\bibitem{beliy2019voxels}
Roman Beliy, Guy Gaziv, Assaf Hoogi, Francesca Strappini, Tal Golan, and Michal Irani.
\newblock From voxels to pixels and back: Self-supervision in natural-image reconstruction from fmri.
\newblock {\em Advances in Neural Information Processing Systems}, 32, 2019.

\bibitem{gaziv2022self}
Guy Gaziv, Roman Beliy, Niv Granot, Assaf Hoogi, Francesca Strappini, Tal Golan, and Michal Irani.
\newblock Self-supervised natural image reconstruction and large-scale semantic classification from brain activity.
\newblock {\em NeuroImage}, 254:119121, 2022.

\bibitem{kay2008identifying}
Kendrick~N Kay, Thomas Naselaris, Ryan~J Prenger, and Jack~L Gallant.
\newblock Identifying natural images from human brain activity.
\newblock {\em Nature}, 452(7185):352--355, 2008.

\bibitem{shen2019deep}
Guohua Shen, Tomoyasu Horikawa, Kei Majima, and Yukiyasu Kamitani.
\newblock Deep image reconstruction from human brain activity.
\newblock {\em PLoS computational biology}, 15(1):e1006633, 2019.

\bibitem{naselaris2009bayesian}
Thomas Naselaris, Ryan~J Prenger, Kendrick~N Kay, Michael Oliver, and Jack~L Gallant.
\newblock Bayesian reconstruction of natural images from human brain activity.
\newblock {\em Neuron}, 63(6):902--915, 2009.

\bibitem{kay2011fmri}
Kendrick~N Kay, Thomas Naselaris, and Jack~L Gallant.
\newblock fmri of human visual areas in response to natural images.
\newblock {\em CRCNS. org}, 2011.

\bibitem{horikawa2020attentionally}
Tomoyasu Horikawa and Yukiyasu Kamitani.
\newblock Attentionally modulated subjective images reconstructed from brain activity.
\newblock {\em bioRxiv}, 2020.

\bibitem{ho2022inter}
Jun~Kai Ho, Tomoyasu Horikawa, Kei Majima, and Yukiyasu Kamitani.
\newblock Inter-individual deep image reconstruction.
\newblock {\em bioRxiv}, pages 2021--12, 2022.

\bibitem{scotti2023reconstructing}
Paul~S Scotti, Atmadeep Banerjee, Jimmie Goode, Stepan Shabalin, Alex Nguyen, Ethan Cohen, Aidan~J Dempster, Nathalie Verlinde, Elad Yundler, David Weisberg, et~al.
\newblock Reconstructing the mind's eye: fmri-to-image with contrastive learning and diffusion priors.
\newblock {\em arXiv preprint arXiv:2305.18274}, 2023.

\bibitem{Deng2009ImageNet:Database}
Jia Deng, Wei Dong, Richard Socher, Li-Jia Li, {Kai Li}, and {Li Fei-Fei}.
\newblock {ImageNet: A large-scale hierarchical image database}.
\newblock In {\em 2009 IEEE Conference on Computer Vision and Pattern Recognition}, pages 248--255. IEEE, 6 2009.

\bibitem{cocodataset}
Tsung{-}Yi Lin, Michael Maire, Serge~J. Belongie, Lubomir~D. Bourdev, Ross~B. Girshick, James Hays, Pietro Perona, Deva Ramanan, Piotr Doll{'{a} }r, and C.~Lawrence Zitnick.
\newblock Microsoft {COCO:} common objects in context.
\newblock {\em CoRR}, abs/1405.0312, 2014.

\bibitem{simonyan2014very}
Karen Simonyan and Andrew Zisserman.
\newblock Very deep convolutional networks for large-scale image recognition.
\newblock {\em arXiv preprint arXiv:1409.1556}, 2014.

\end{thebibliography}
\bibliographystyle{unsrt}

\clearpage
\part*{Supplementary Material}

\setcounter{page}{1}

\setcounter{figure}{0}
\setcounter{table}{0}
\setcounter{section}{0}
\renewcommand{\thefigure}{S\arabic{figure}}
\renewcommand{\thetable}{S\arabic{table}}
\renewcommand{\thesection}{S\arabic{section}}

\section{Analyzing Encoder Enhancement}
\label{sec:ablation}

\begin{figure*}[!ht]
     \centering
     \begin{subfigure}{\textwidth}
         \centering
        \includegraphics[width=\textwidth]{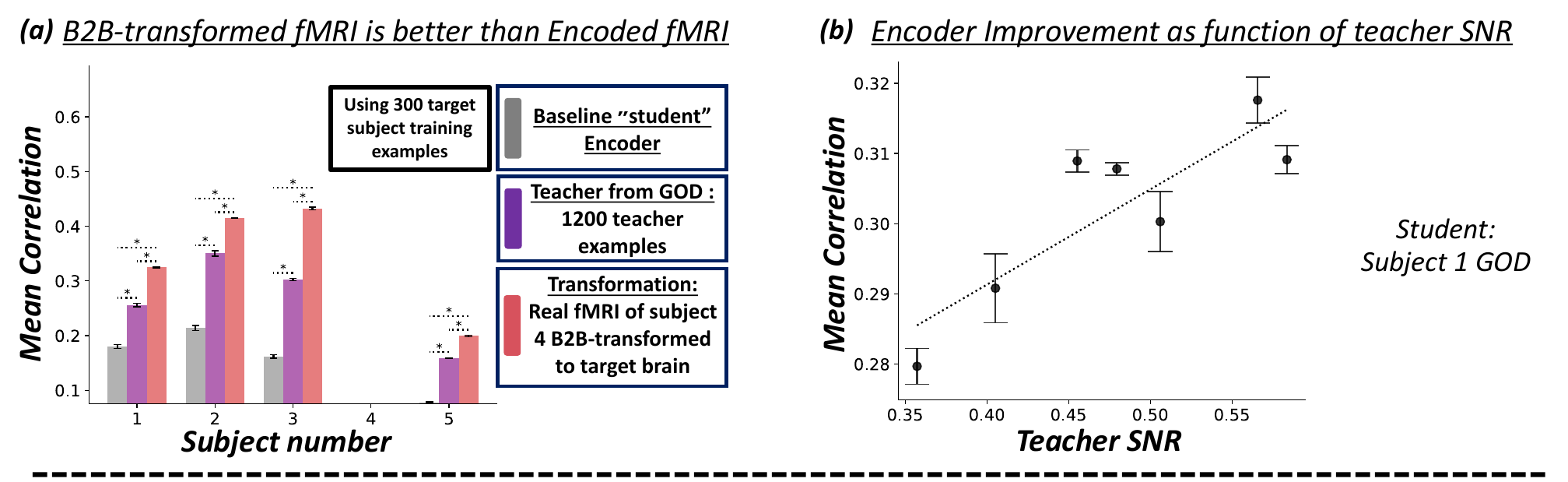}
     \end{subfigure}
     \centering
     \begin{subfigure}{\textwidth}
         \centering
        \includegraphics[width=\textwidth]{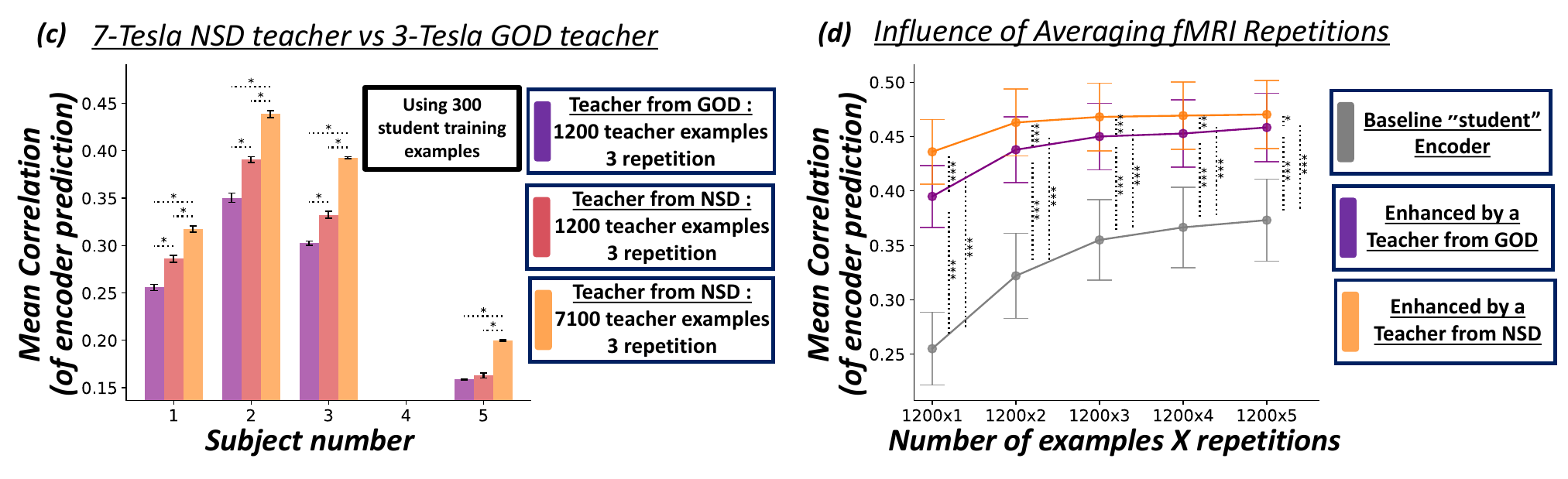}
     \end{subfigure}
     
\caption{{\textbf{Encoder Enhancement Analyzing:}
\textbf{(a)}~The figure demonstrates that a real fMRI scan from Subject 4, after being B2B-transformed into the space of target subjects (Subjects 1, 2, 3, and 5), outperforms the Image-to-fMRI encoding of the target subjects' Baseline encoders trained on 300 examples. Grey bars represent the Baseline-encoder performance, while red bars show the Pearson correlation between the B2B-transformed fMRI and the target subjects' real fMRI scans, evaluated on GOD test data. This highlights the potential of B2B-transformed fMRI to enhance encoder performance, particularly for subjects with limited training data.
\textbf{(b)}~Teacher quality analysis: the correlation between teacher voxels mean SNR and encoding performance, shown for GOD Subject 1 using NSD subjects as teachers (mean and SEM across three repetitions). Higher-quality teacher data leads to better encoding performance showed by the postive correlation between the teacher SNR with the average mean correlations over repetitions (\(r=0.85\), \(R^2=0.735\), \(p=0.006\)).
\textbf{(c)}~Comparison of GOD student encoder performance with 1200 training examples enhanced by GOD or NSD teachers. Models trained with NSD teachers consistently outperform those trained with GOD teachers, with further improvement observed when all NSD examples are utilized.
\textbf{(d)}~Impact of repetitions: using more repetitions per example in the GOD dataset improves encoding performance, as shown for GOD subjects. Even without additional repetitions, our method achieves notable performance gains.}}
\vspace{0.5cm}
\label{Figure:Ablation}
\end{figure*}

\vspace{0.2cm}
\noindent
{In this section, we analyze various factors that contribute to improving Image-to-fMRI encoder performance and offer deeper insights into the effectiveness of our method. \cref{Figure:Ablation} consolidates four plots that dissect different aspects of our approach.}
\\

\noindent
{We first show that a \emph{real} fMRI scan, when B2B-transformed from a source-subject to a target-subject, exceeds the quality of Image-to-fMRI encoding of the target-subject (and is particularly powerful if the target-subject has only a small number of examples to train its encoder). 
This assumption is validated in \cref{Figure:Ablation}.a, where the source subject was Subject4 of the GOD datasdet, and the 4 other GOD subjects (Subject$_i$, i=1,2,3,5) served as target subjects. The Image-to-fMRI encoder of the 4 target subjects were trained using 300 (25\%) of their training examples. The figure shows that a \emph{real} fMRI of Subject4, when B2B-transformed to a target subject, outperforms the emph{encoded}-fMRI obtained from the target-subject's Baseline-encoder. 
The grey bars represent the performance of the Baseline-encoder, while the red bars show the mean Pearson correlation between the real fMRI of GOD Subject4 after being B2B-transformed into the space of the target-subject. The correlations are measured with respect to the real fMRI scan of the target-subject (evaluated on the GOD test-data). 
This indicates that the B2B-transformed fMRI can indeed be used to boost encoder performance of another subject (especially when that subject has few training data).} 
\\

\begin{figure*}[!hb]
    \centering
    \begin{subfigure}{\textwidth}
        \includegraphics[width=\textwidth]{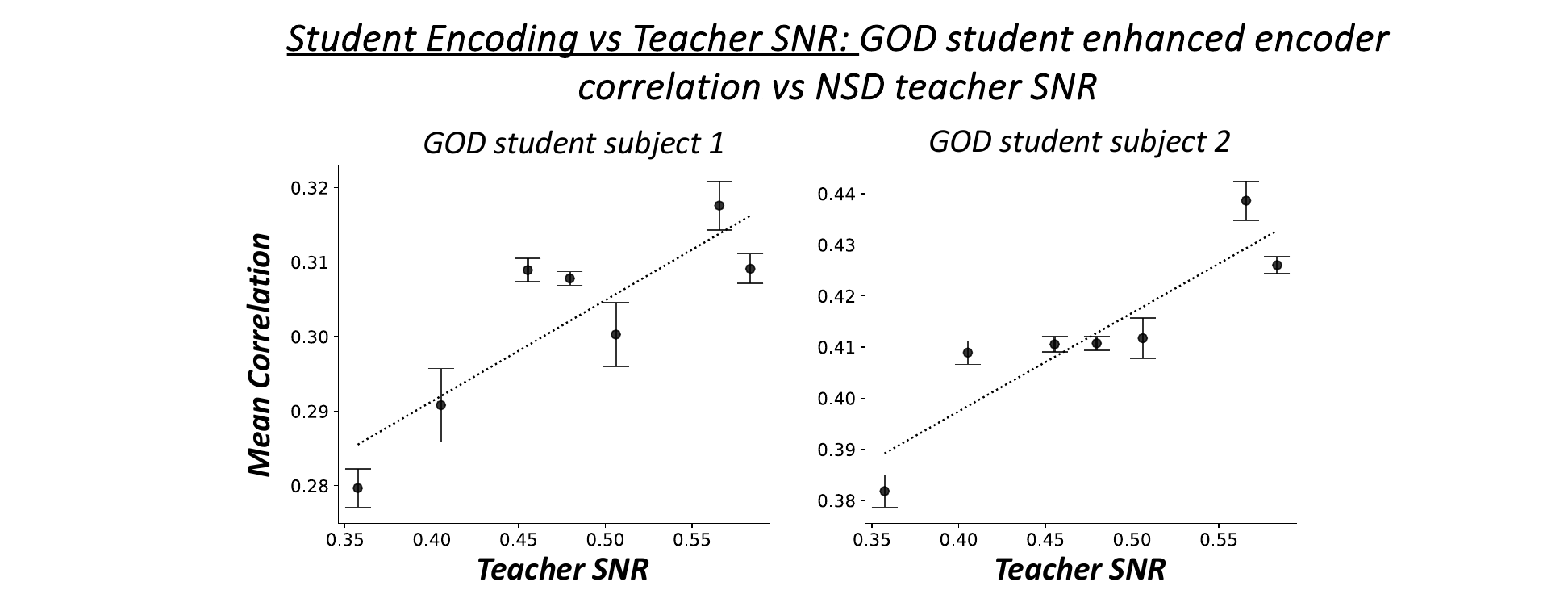} 
    \end{subfigure}
    \begin{subfigure}{\textwidth}
        \includegraphics[width=\textwidth]{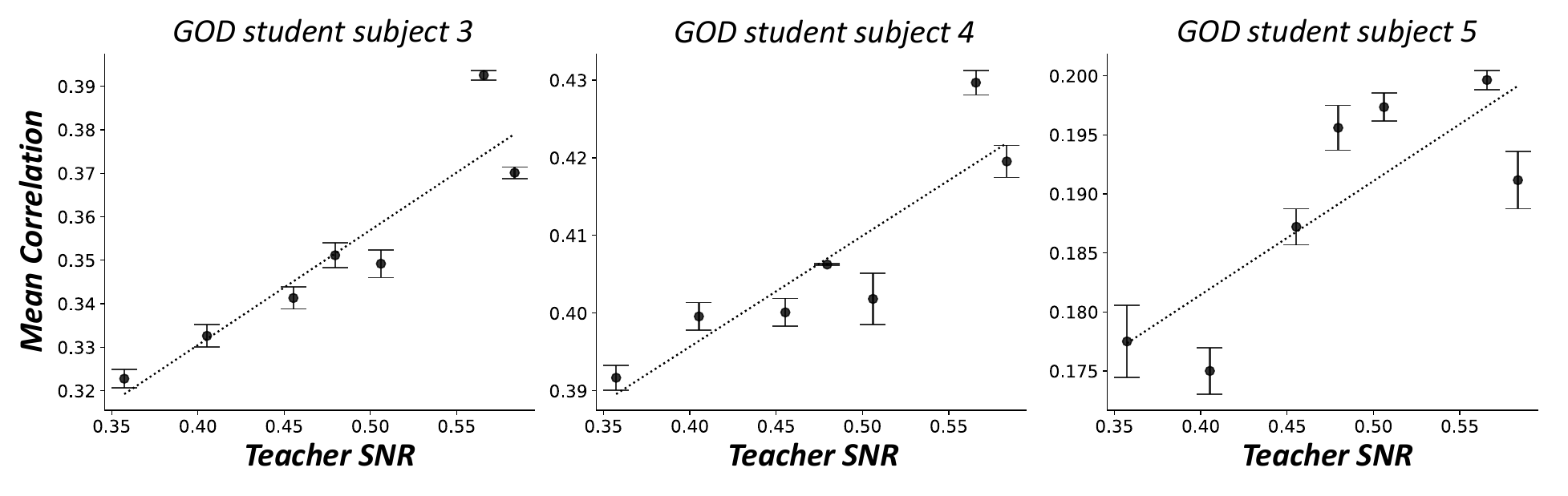} 
    \end{subfigure}
    \caption{\textbf{Teacher quality analysis:} the correlation between teacher SNR and encoding performance, shown for GOD Subjects using NSD subjects as teachers (mean and SEM across three repetitions). Higher-quality teacher data leads to better encoding improvements.}
    \label{SM_Figure:SNR}
\end{figure*}

\noindent
{The quality of the ``teacher" data is another critical factor in improving encoder performance. In~\cref{Figure:Ablation}.b, we demonstrate that a subject with higher-quality data serves as a more effective teacher. For this analysis, we still used GOD subjects as students, but this time with NSD subjects as teachers (we used 7 NSD subjects as teachers, excluding one outlier with a significant gap between data quality and encoding performance). We trained each GOD subject with one (out of 7) NSD subject as a teacher on 300 examples. For each GOD subject, we obtained seven encoding correlations, corresponding to the seven NSD teachers. We perform this experiment for each of the 3 repeats and average them. The figure shows the mean and SEM for GOD Subject 1 (see~\cref{SM_Figure:SNR} for the remaining subjects). A clear trend emerges, with a correlation between teacher SNR and encoding performance.}
\\

\noindent
{In~\cref{Figure:Ablation}.c, we first analyze the benefits of using a 7-Tesla NSD teacher versus {a 3-Tesla} GOD teacher, with the same number of \emph{teacher} training examples (purple versus orange bars). The results show the performance of each GOD subject trained with 300 examples, enhanced via a teacher subject. We compare using Subject4 from GOD (purple) and Subject1 from NSD (red) as teachers, ensuring a fair comparison by using teacher data with the same number of available examples (1200 training pairs and three repetitions per example). The plot reveals that using NSD as the teacher consistently outperforms the GOD teacher across all four GOD student subjects. Additionally, when a large number (7100) of the NSD teacher training examples are used (orange), there is a noticeable performance boost, highlighting the importance of both high-quality teacher fMRI data and a larger number of examples in improving encoder performance.}
\\

\begin{figure*}[!hbt]
    \centering
    \begin{subfigure}{\textwidth}
        \includegraphics[width=\textwidth]{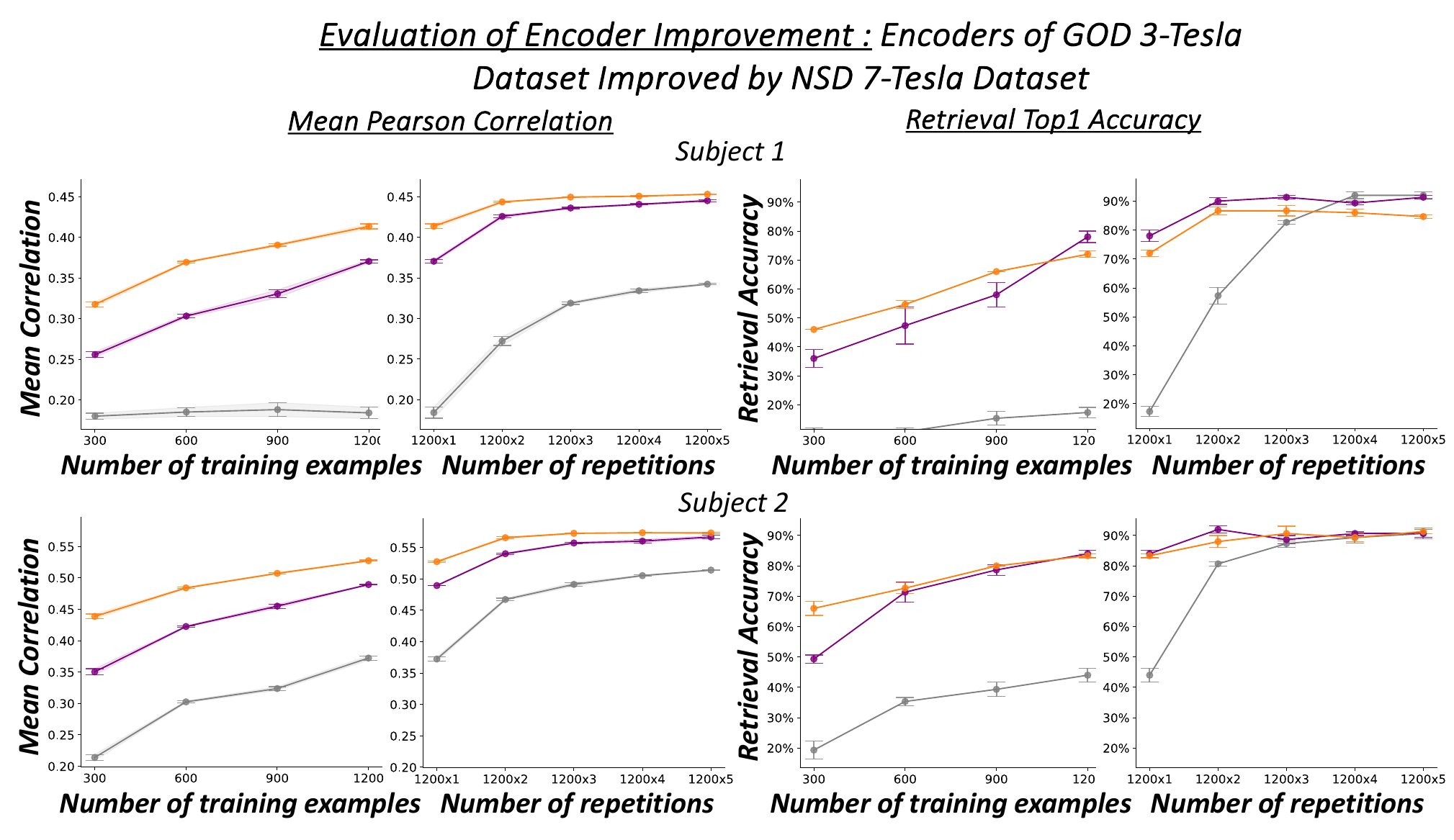} 
    \end{subfigure}
    \begin{subfigure}{\textwidth}
        \includegraphics[width=\textwidth]{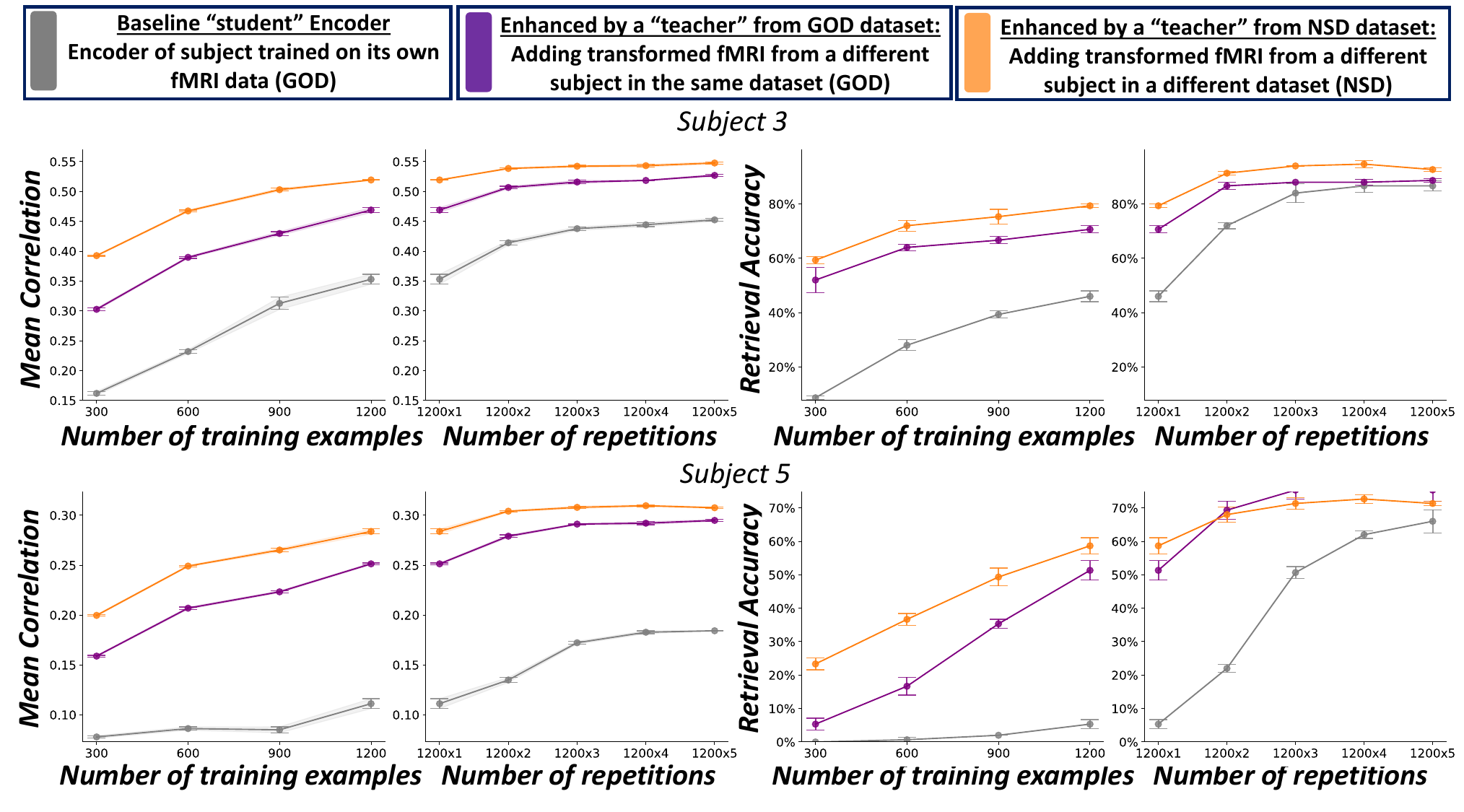} 
    \end{subfigure}
    
    \caption{\textbf{Improving encoding of a subject from GOD 3-Tesla dataset using another high-quality subject:}
     This figures compares the quality of the ``student" baseline encoder (which is trained solely on the student's own fMRI data), to the student encoder obtained when incorporating also data from the ``teacher" during training. In each plot, the grey line represents the student baseline encoder model, the purple line symbolizes the encoder improvement obtained when using a ``teacher" from the same dataset (e.g., Subject4 is the highest-quality subject in the GOD dataset). The orange line corresponds to the encoder enhancement achieved by utilizing a superior subject from another dataset (NSD), which has more scanned examples and higher fMRI resolution (7-Tesla machine). The quality of the resulting encoders is assessed through the mean Pearson correlation of all predicted fMRI voxels. The ``number of training examples" refers to the number of the student's own image-fMRI pairs of examples used for training the ``student" encoder, whereas the ``teacher" subject (drawn from either GOD or NSD dataset) has been trained using all its own available examples and remains unchanged during the training of the student encoder. All plots present specific subject encoding results quantified using the mean Pearson correlation of all voxels across different models, based on the number of training examples. 
    }
    \label{SM_Figure:Encoder_GOD}
\end{figure*}

\noindent
{Finally, we evaluate the impact of increasing the number of repetitions in the GOD dataset. The updated GOD dataset version includes five repetitions per example. As shown in~\cref{Figure:Ablation}.d and in~\cref{SM_Figure:Encoder_GOD}, more repetitions lead to improved encoder performance. This is expected, as averaging more repetitions reduces noise in the fMRI, thus improving its signal-to-noise ratio (SNR). Interestingly, even without additional repetitions, our B2B method for enhancing encoders produces high performance results, suggesting that our approach could serve as an alternative to acquiring more data or repetitions. }

\vspace{2cm}
\begin{table}[!h]
\centering
\begin{tabular}{|c|c|c|c|c|}
\hline
\textbf{Subject} & \textbf{Baseline Encoder} & \textbf{+ Shuffled Data} & \textbf{+ Random Data} & \textbf{+ Duplicated Data} \\ \hline
\textbf{1} & 0.1830 & 0.1416 & 0.1283 & 0.1725 \\
\textbf{2} & 0.2152 & 0.1701 & 0.2047 & 0.2153 \\ 
\textbf{3} & 0.1590 & 0.1131 & 0.1280 & 0.1626 \\ 
\textbf{4} & 0.2515 & 0.1947 & 0.2033 & 0.2478 \\ 
\textbf{5} & 0.0779 & 0.0638 & 0.0670 & 0.0813 \\ \hline
\textbf{Average} & \textbf{0.1773} & \textbf{0.1289} & \textbf{0.1462} & \textbf{0.1759} \\ \hline
\end{tabular}
\caption{{\textbf{Impact of Augmented Data on Encoder Performance:} This experiment evaluates whether the improvement seen in the encoder’s performance, as demonstrated in the paper, is due to the addition of meaningful data from a teacher subject or simply the result of having more training data. To explore this, we present results for three different augmentation methods used to add 300 additional examples to the baseline model, which was initially trained with 300 original examples. These methods include: (i) shuffling voxel values within the original fMRI data to generate synthetic examples, (ii) creating random fMRI data with the same distribution as the measured data, and (iii) duplicating the original 300 examples. The results show that these simple augmentation methods fail to significantly improve performance, underscoring the importance of meaningful teacher-derived data for boosting encoder performance.}}
\label{table:augmented_data_ablation}
\end{table}

\newpage
\begin{figure*}[!hbt]
    \centering
    \begin{subfigure}{\textwidth}
        \includegraphics[width=\textwidth]{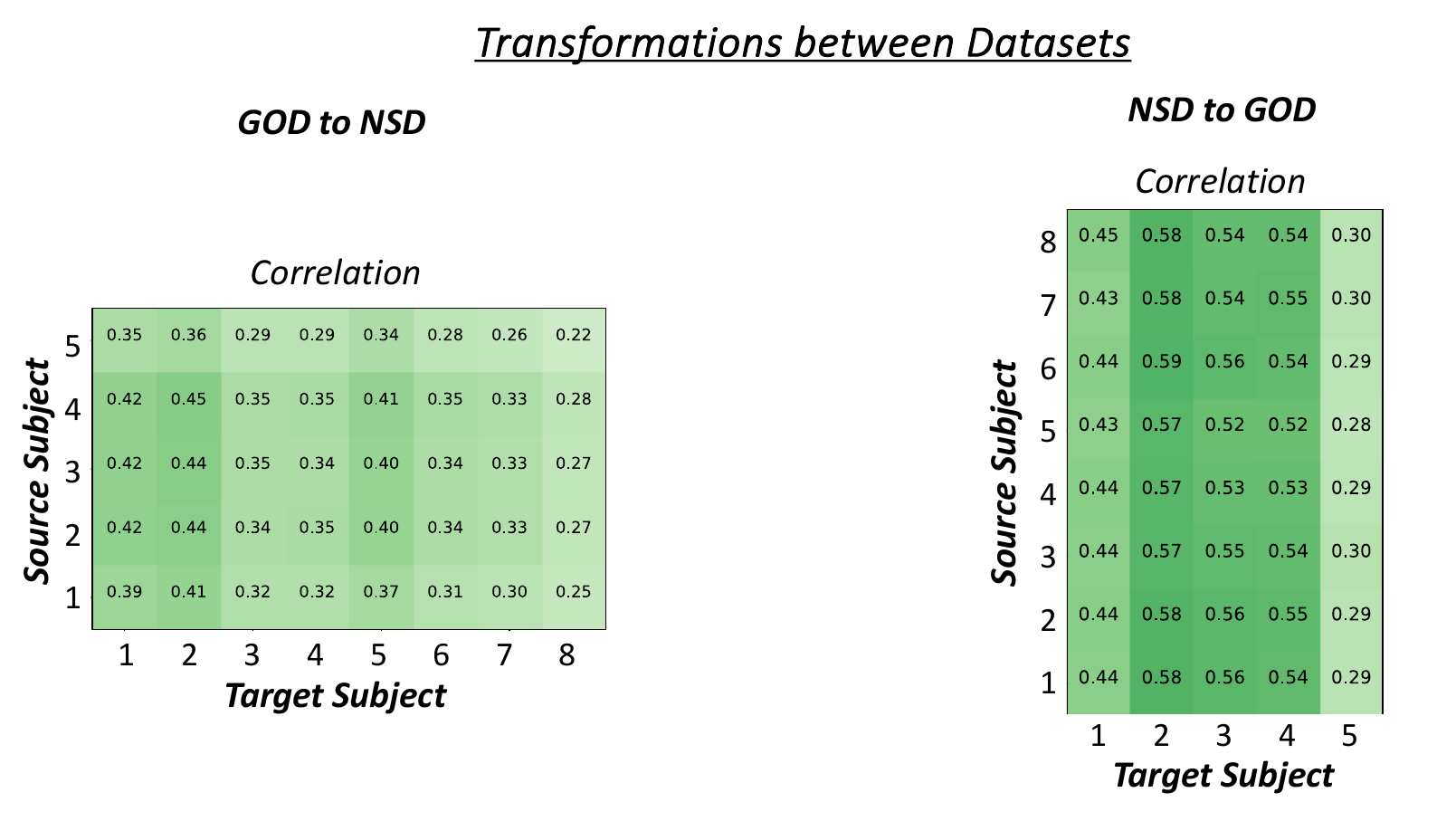} 
    \end{subfigure}
    \caption{\textbf{Transformation between NSD and GOD datasets:} This figure presents two heatmaps showing the mean Pearson correlation for all subject pairs: one for transformations from GOD to NSD and the other from NSD to GOD. As there is no shared data between the datasets, we generated corresponding fMRIs for the target dataset's test set using the encoder trained on the source dataset. These encoded fMRIs were then transformed to match the target subject and evaluated by comparing them to the real measured test fMRIs.}
\label{SM_Figure:Transformation_across_datasets}
\end{figure*}

\begin{figure*}[!hbt]
    \centering
    \begin{subfigure}{\textwidth}
        \includegraphics[width=\textwidth]{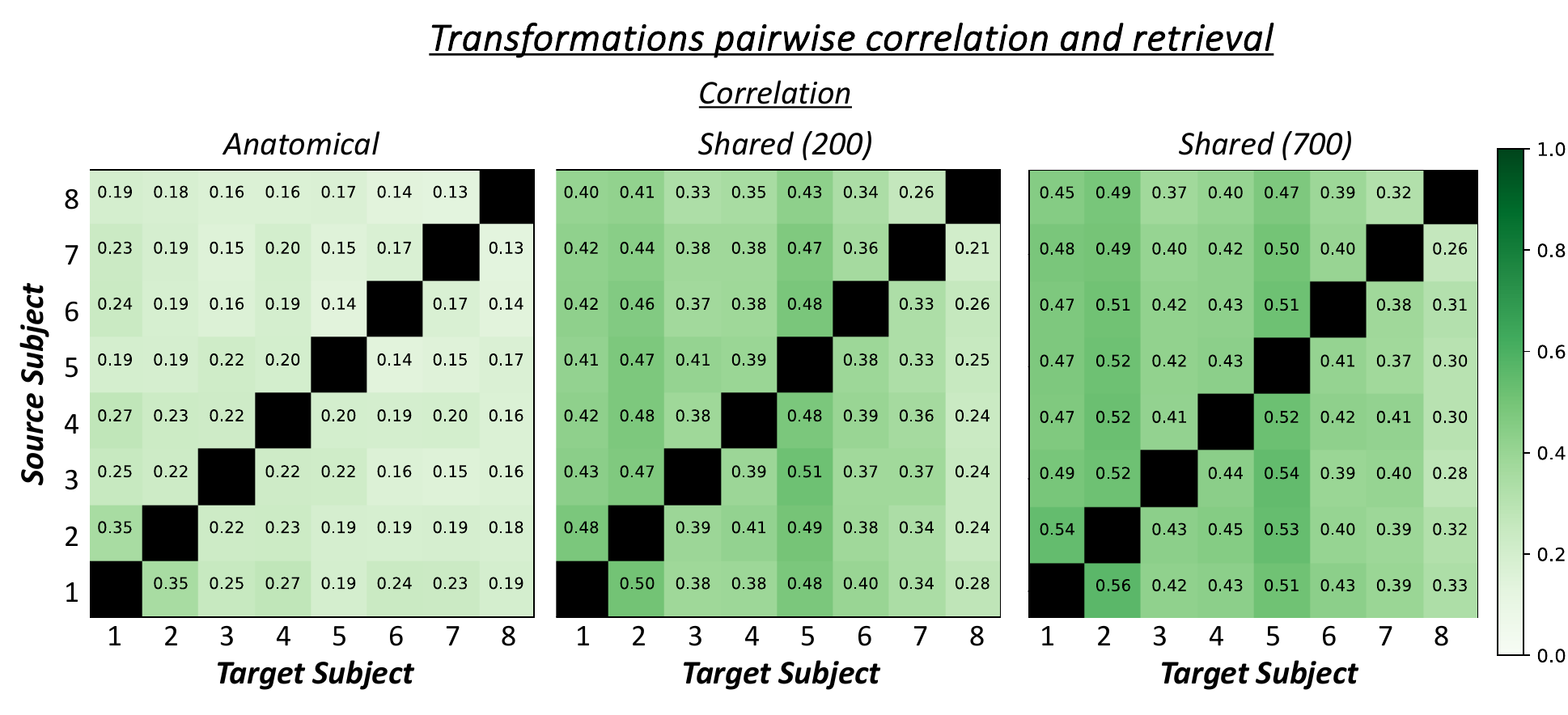} 
    \end{subfigure}
    \begin{subfigure}{\textwidth}
        \includegraphics[width=\textwidth]{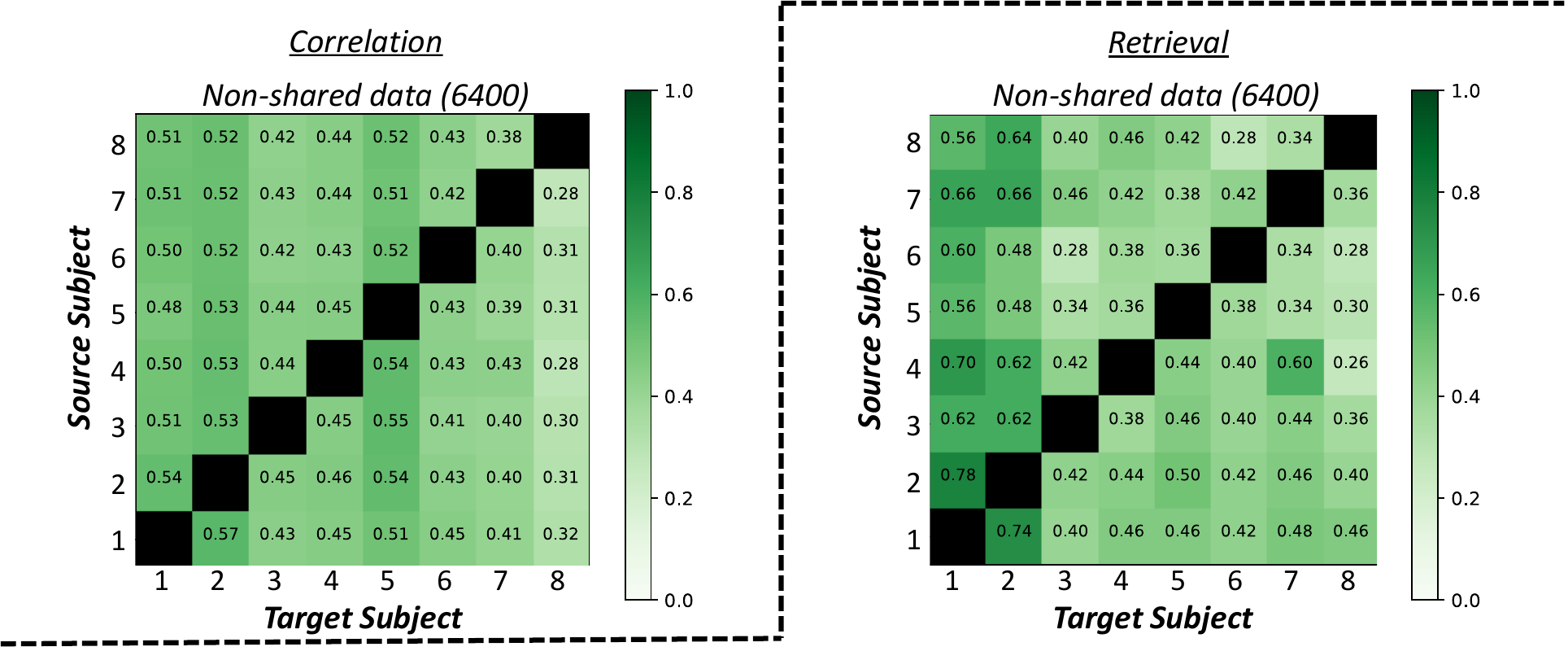} 
    \end{subfigure}
    \begin{subfigure}{\textwidth}
        \includegraphics[width=\textwidth]{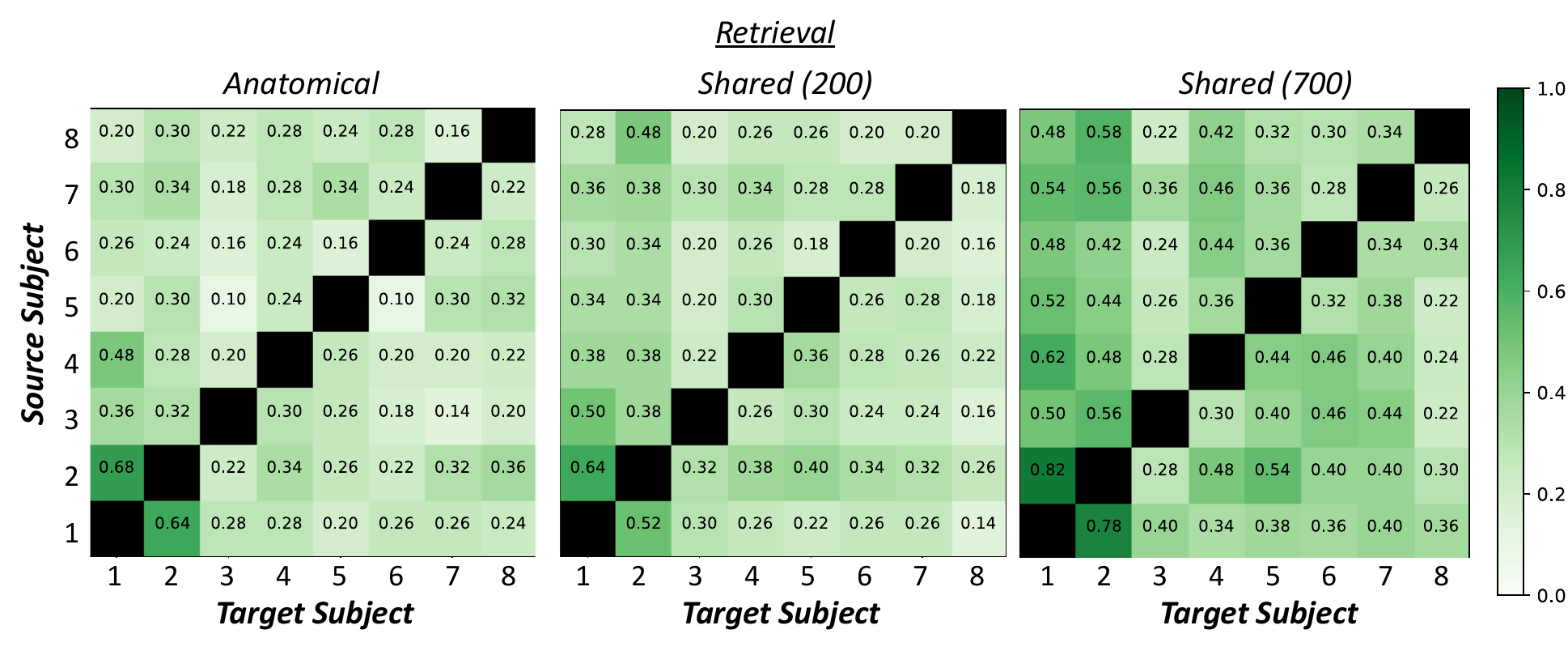} 
    \end{subfigure}
    \caption{\textbf{Pairwise Correlation and Retrieval for Transformations:} The figure presents heatmaps for all NSD subject pairs, showing both correlation and retrieval accuracy for four different transformation methods: (i) anatomical mappings, (ii) transformations using 200 shared examples, (iii) transformations using 700 shared examples, and (iv) our method, which leverages all available non-shared data (6400 examples) and external images. 
    }
    \label{SM_Figure:Transformation_hetmaps_NSD}
\end{figure*}

\begin{figure}[ht]     
     \begin{subfigure}{0.45\textwidth}
         \includegraphics[width=\textwidth]{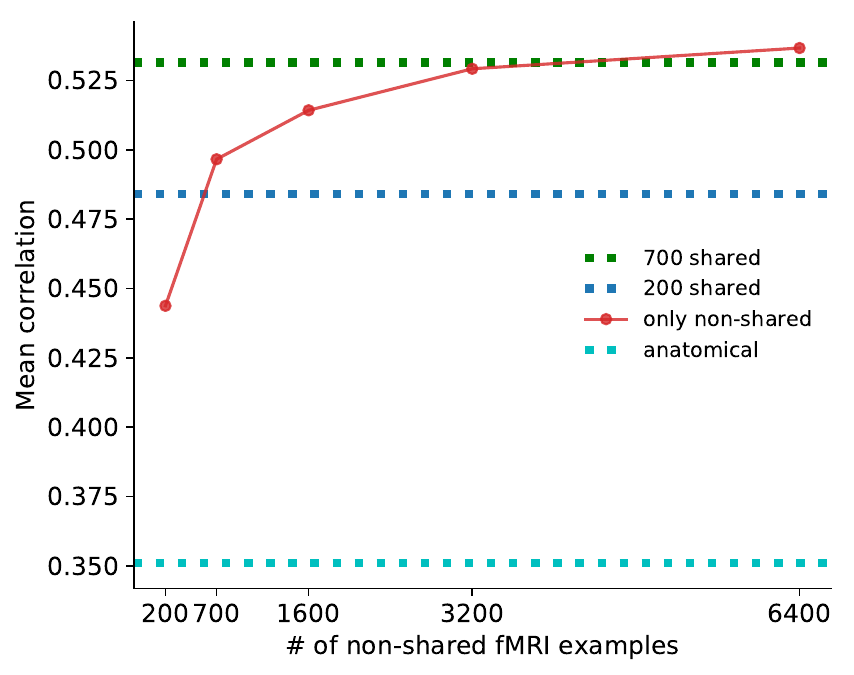} 
         \caption{Transformation of Subject1 to Subject2}
     \end{subfigure}
      \hfill
     \begin{subfigure}{0.45\textwidth}
         \includegraphics[width=\textwidth]{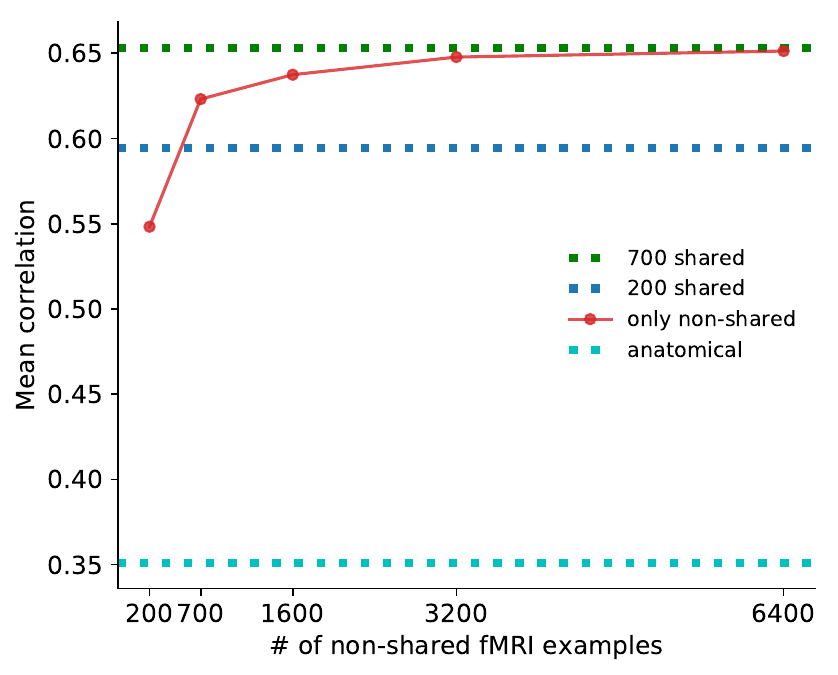} 
         \caption{Transformation of Subject2 to Subject1}
     \end{subfigure}
     \hfill
     \begin{subfigure}{0.45\textwidth}
         \includegraphics[width=\textwidth]{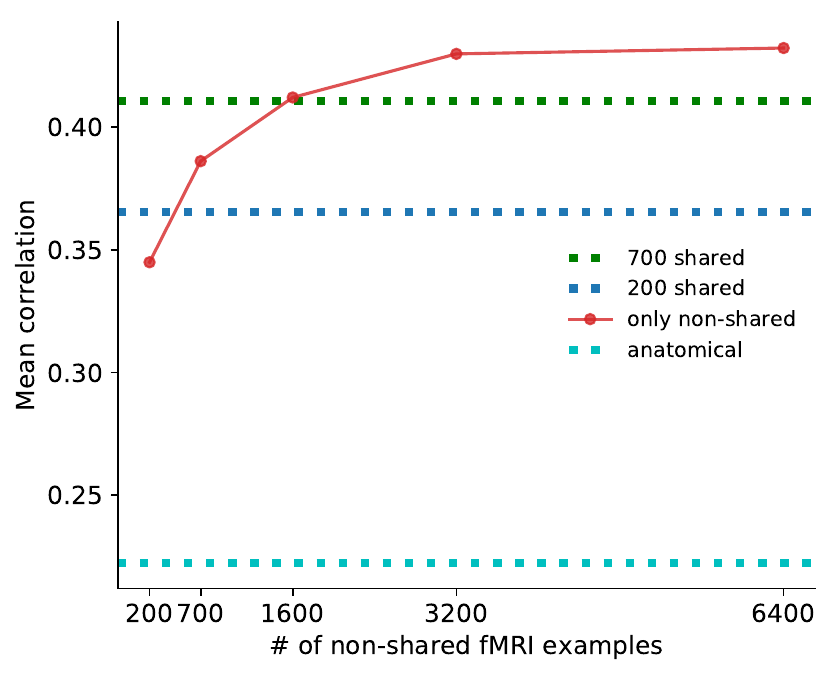}
         \caption{ Transformation of Subject3 to Subject4}
     \end{subfigure}
     \hfill
     \begin{subfigure}{0.45\textwidth}
         \includegraphics[width=\textwidth]{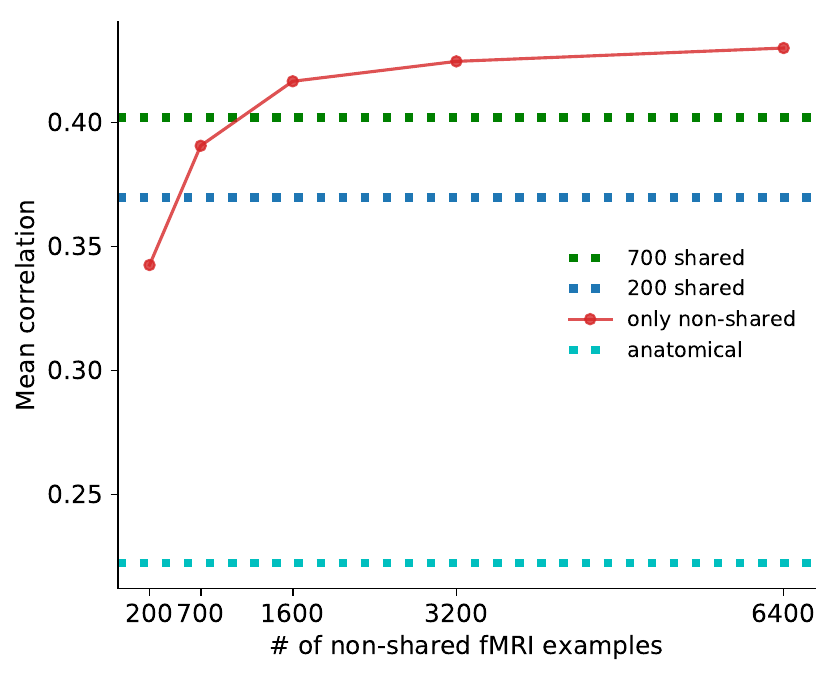}
         \caption{Transformation of Subject4 to Subject3}
     \end{subfigure}
      \caption{\textbf{Evaluating brain-to-brain transformation with no shared data}:
        In this figure, we conduct a quantitative assessment by calculating the mean Pearson’s r correlation for different transformations within the NSD dataset across various transformation models. We compare the resulting transformations when trained only with non-shared data (solid red line) to three other transformations: anatomical mapping to a common brain space (cyan dotted line), transformation trained with only 200 shared examples (dotted blue line), and transformation trained with 700 shared examples (dotted grey line). The remaining 300 ``shared data" serve as ``test data" for assessing and comparing the quality of all the learned transformations. The x-axis of the graph represents the number of non-shared examples available for training, where our transformation model is the only one that can use non-shared examples.
        \textbf{(a)}: A single example of the Pearson's r results for the transformation between Subject1 and Subject2 in the NSD dataset. 
        \textbf{(b)}: Same as 'a', between Subject2 and Subject1.  
        \textbf{(c)}: Same as 'a', between Subject3 and Subject4. 
        \textbf{(d)}: Same as 'a', between Subject4 and Subject3. }
        \label{SM_Figure:Tranformation_singles}
\end{figure}

\begin{figure*}     
     \begin{subfigure}{0.45\textwidth}
         \includegraphics[width=\textwidth]{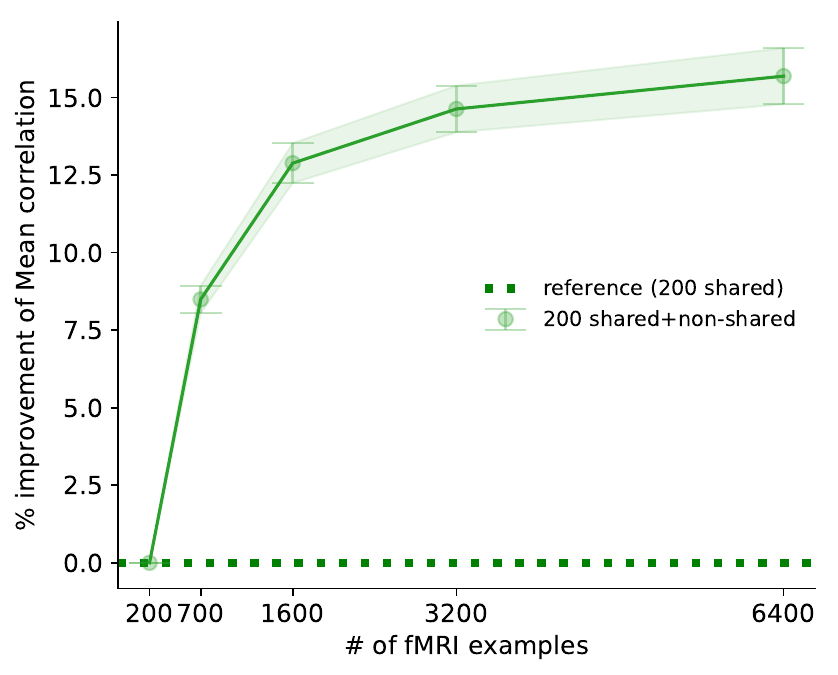} 
         \caption{Mean of all transformations}
     \end{subfigure}
      \hfill
     \begin{subfigure}{0.45\textwidth}
         \includegraphics[width=\textwidth]{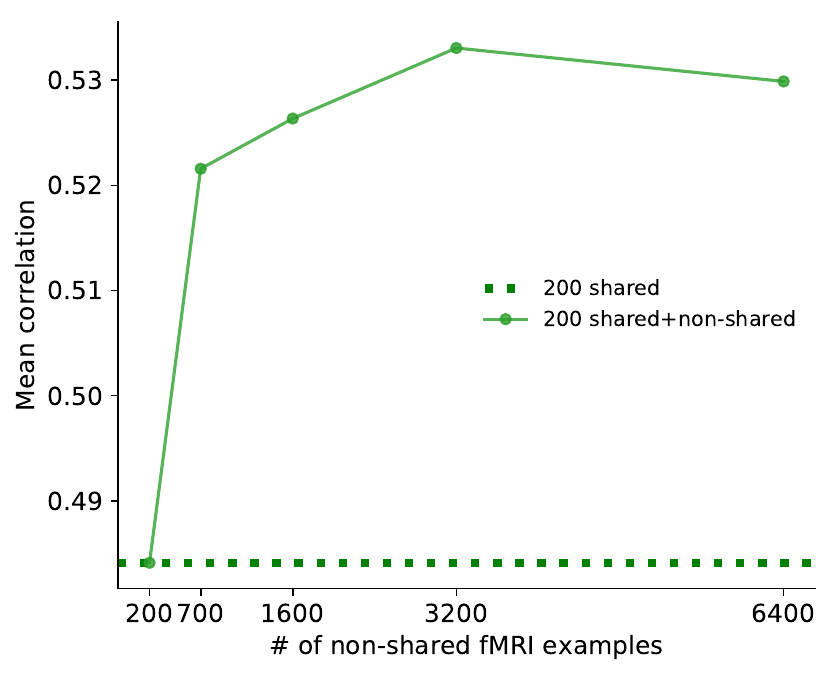} 
         \caption{Transformation Subject1 to Subject2}
     \end{subfigure}
     \hfill
     \begin{subfigure}{0.45\textwidth}
         \includegraphics[width=\textwidth]{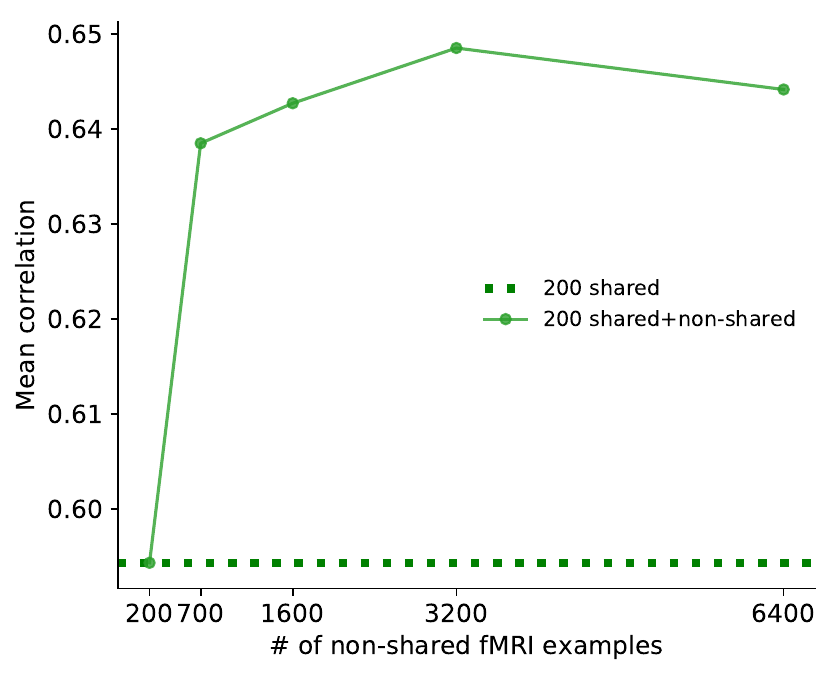}
         \caption{ Transformation Subject2 to Subject1}
     \end{subfigure}
     \hfill
     \begin{subfigure}{0.45\textwidth}
         \includegraphics[width=\textwidth]{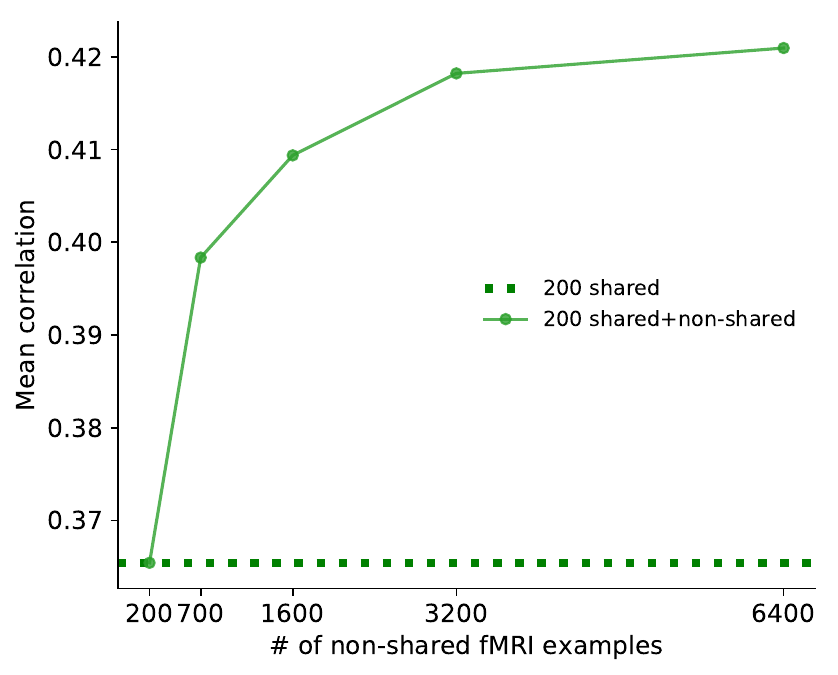}
         \caption{Transformation Subject3 to Subject4}
     \end{subfigure}
    \caption{\textbf{Using non shared data alongside shared data to improve transformations}:
    Across all figures, we make a comparison between a transformation trained with only 200 shared examples (shown as a dotted {green} line) and a transformation trained using our method, which includes non-shared examples in addition to the 200 shared examples (shown as a green line). The x-axis on each figure represents the quantity of non-shared examples utilized for training. 
    \textbf{(a)}: Percentage improvement in Pearson’s r for models using varying amounts of non-shared data (green line) compared to the transformation trained with 200 shared examples (dotted {green} line). The error bars represent the SEM over the pairs.
    \textbf{(b)}: A single example of the Pearson's r results for the transformation between Subject1 and Subject2 in the NSD dataset. 
    \textbf{(c)}: Same as 'b', between Subject2 and Subject1.   
    \textbf{(d)}: Same as 'b', between Subject3 and Subject4. }
    \label{S2}
\end{figure*}

\begin{figure*}     
     \begin{subfigure}{0.45\textwidth}
         \includegraphics[width=\textwidth]{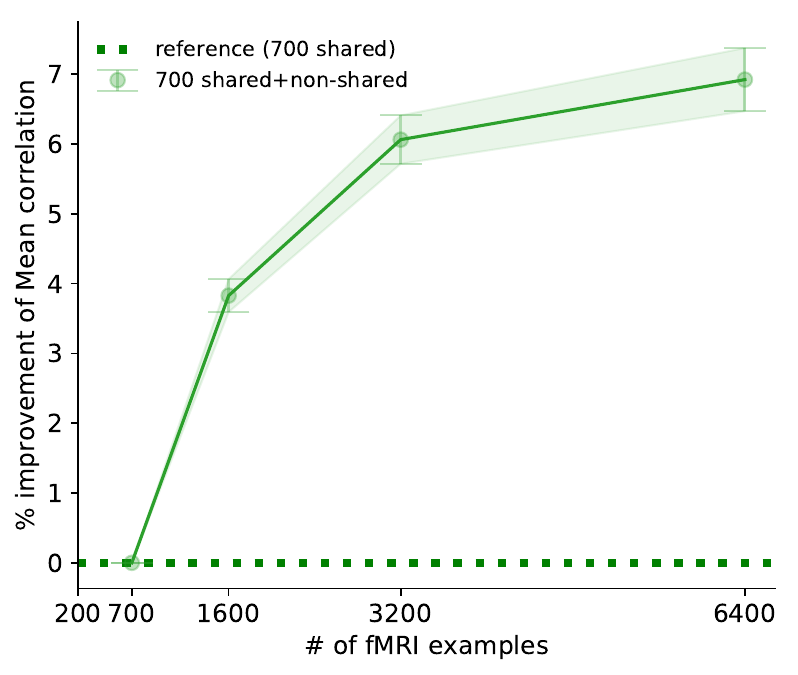} 
         \caption{Mean of all transformations}
     \end{subfigure}
      \hfill
     \begin{subfigure}{0.45\textwidth}
         \includegraphics[width=\textwidth]{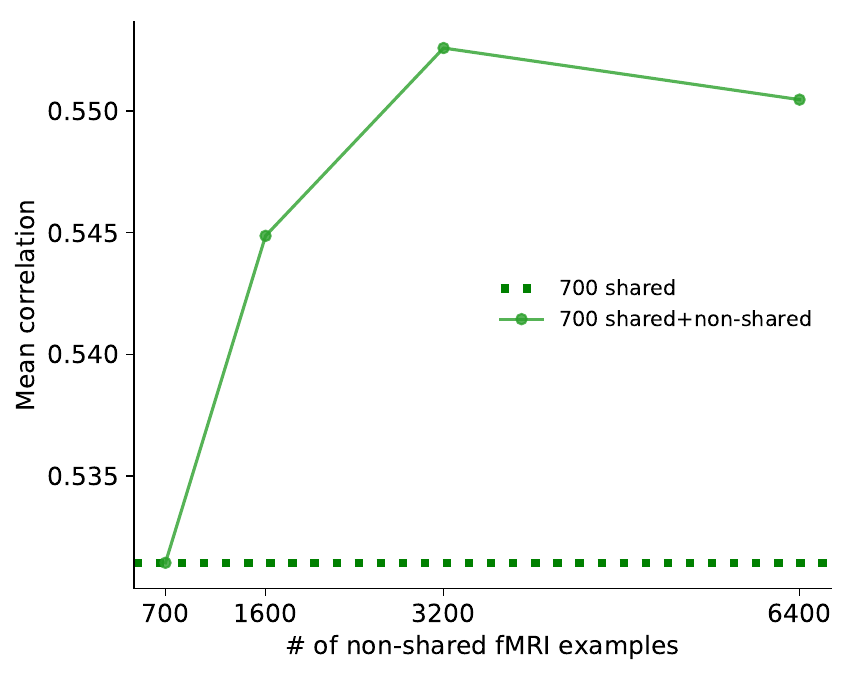} 
         \caption{Transformation Subject1 to Subject2}
     \end{subfigure}
     \hfill
     \begin{subfigure}{0.45\textwidth}
         \includegraphics[width=\textwidth]{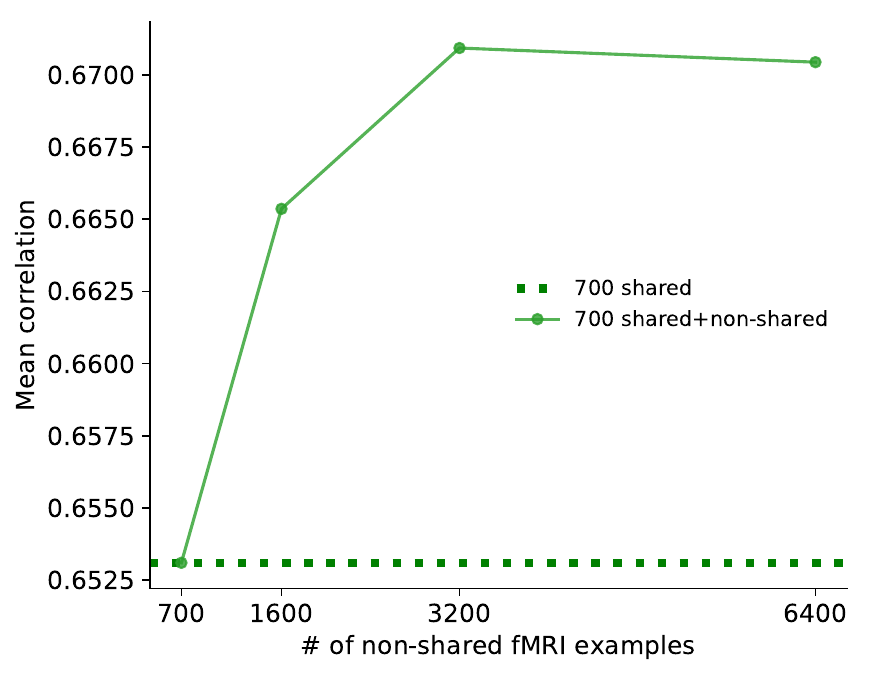}
         \caption{ Transformation Subject2 to Subject1}
     \end{subfigure}
     \hfill
     \begin{subfigure}{0.45\textwidth}
         \includegraphics[width=\textwidth]{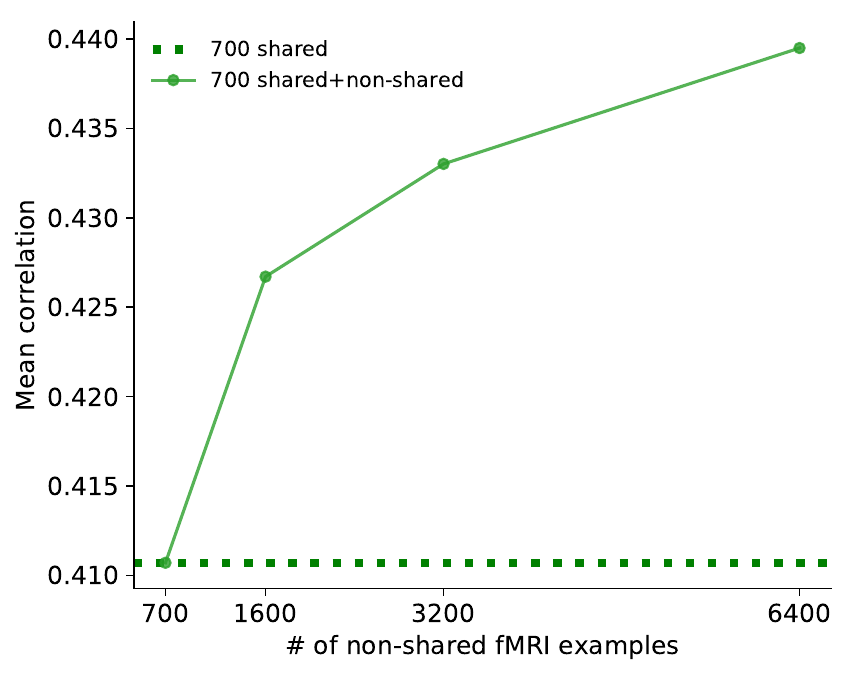}
         \caption{Transformation Subject3 to Subject4}
     \end{subfigure}
    \caption{\textbf{Using non shared data alongside shared data to improve transformations:}
    Across all figures, we make a comparison between a transformation trained with only 700 shared examples (shown as a dotted {green} line) and a transformation trained using our method, which includes non-shared examples in addition to the 700 shared examples (shown as a green line). The x-axis on each figure represents the quantity of non-shared examples utilized for training. 
    \textbf{(a)}: Percentage improvement in Pearson’s r for models using varying amounts of non-shared data (green line) compared to the transformation trained with 700 shared examples (dotted {green} line). The error bars represent the SEM over the pairs. 
    \textbf{(b)}: A single example of the Pearson's r results for the transformation between Subject1 and Subject2 in the NSD dataset. 
    \textbf{(c)}: Same as 'b', between Subject2 and Subject1.   
    \textbf{(d)}: Same as 'b', between Subject3 and Subject4. }
    \label{S3}
\end{figure*}

\begin{figure*}[!hbt]
    \centering
    \begin{tabular}{cc}
         \includegraphics[width=0.95\columnwidth]{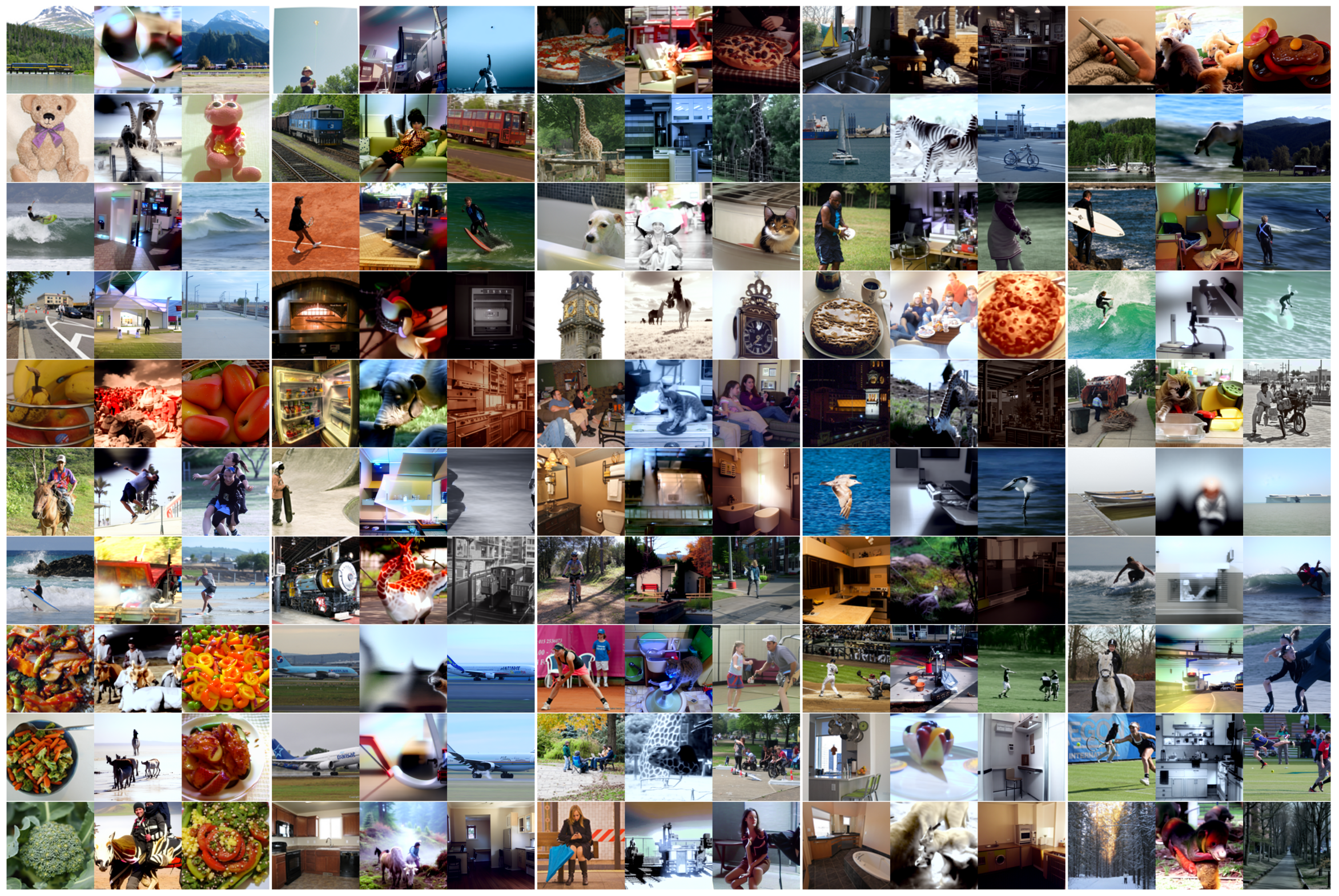} &
    \end{tabular}
    \caption{\textbf{Image Reconstruction from Transformed fMRI of Subject1 to Subject2}:
    Subject1's fMRI data was transformed to align with the fMRI space of Subject2. This transformed data was then decoded into an image using a pre-trained decoder specific to Subject2. The left column displays the original image presented to Subject1, while the second column illustrates the decoded images following anatomical transformation. The subsequent column presents the decoded image after employing our method, trained exclusively with non-shared examples.
    }
    \label{S4}
\end{figure*}

\begin{figure*}[!hbt]
    \centering
    \begin{tabular}{cc}
         \includegraphics[width=0.95\columnwidth]{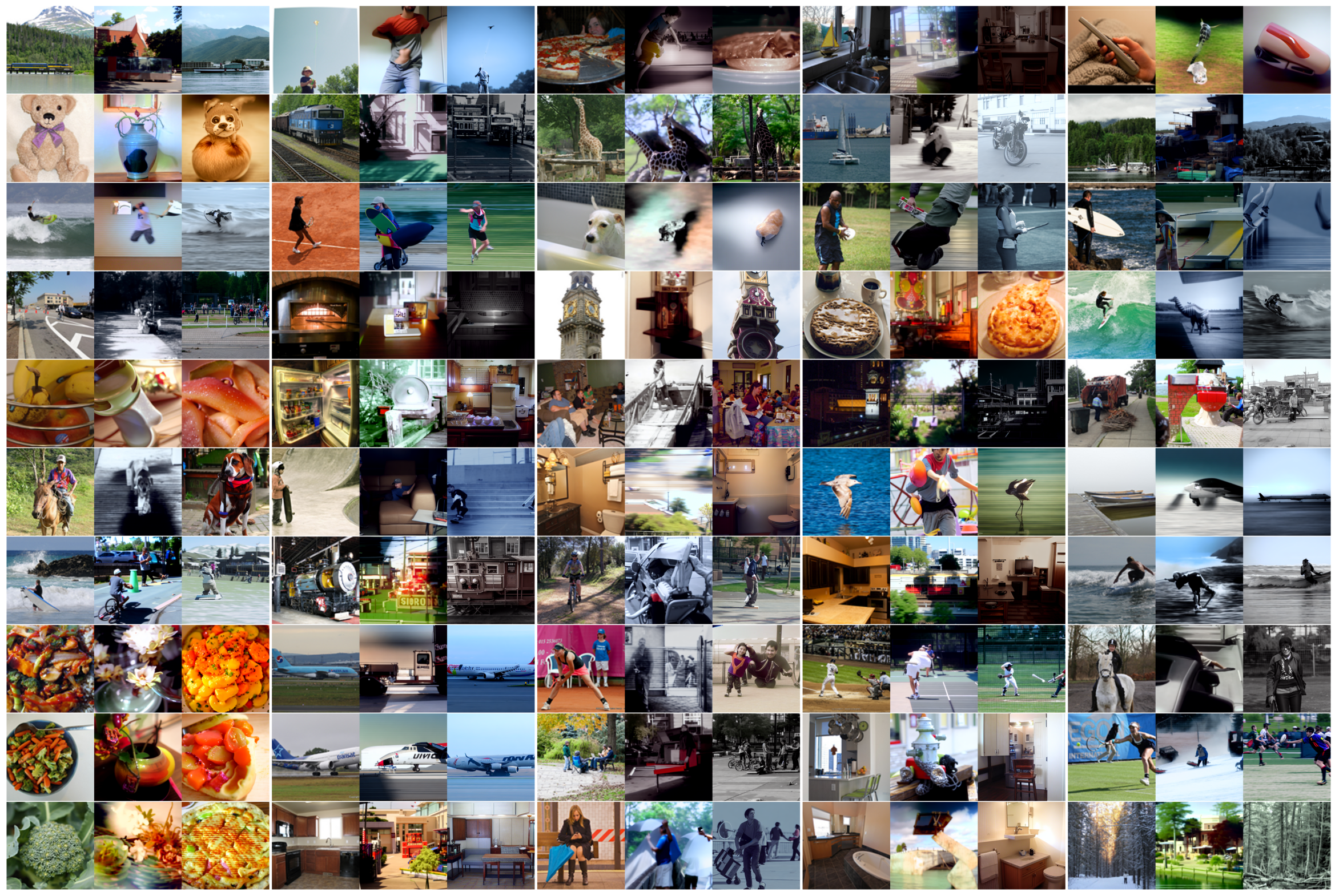} &
    \end{tabular}
    \caption{\textbf{Image Reconstruction from Transformed fMRI of Subject2 to Subject1}:
    Subject2's fMRI data was transformed to align with the fMRI space of Subject1. This transformed data was then decoded into an image using a pre-trained decoder specific to Subject1. The left column displays the original image presented to Subject2, while the second column illustrates the decoded images following anatomical transformation. The subsequent column presents the decoded image after employing our method, trained exclusively with non-shared examples.
    }
    \label{S5}
\end{figure*}

\begin{figure*}[!hbt]
    \centering
    \begin{tabular}{cc}
         \includegraphics[width=0.95\columnwidth]{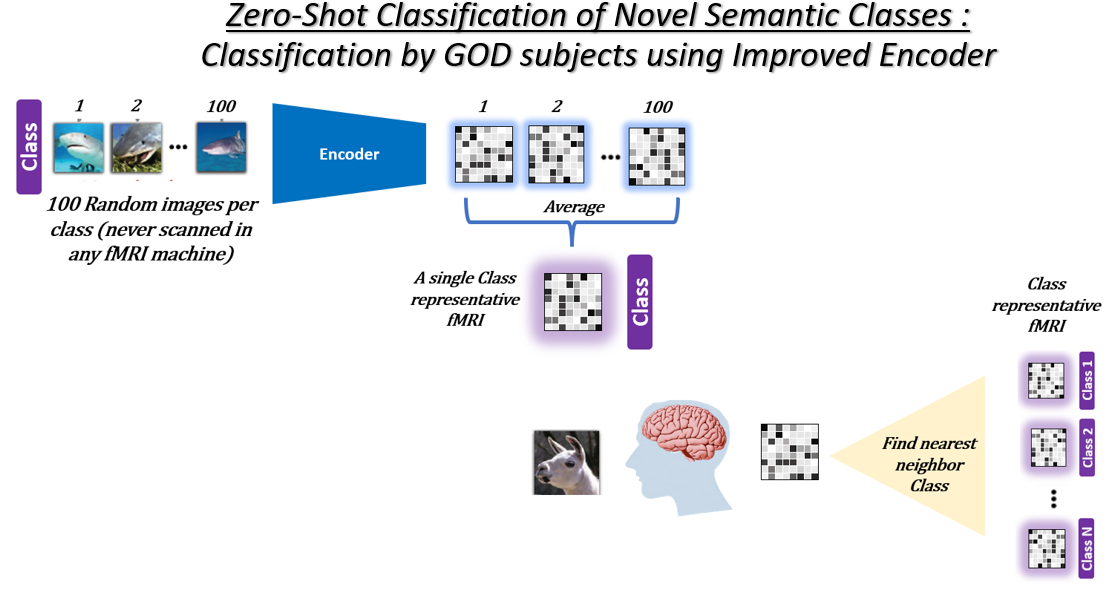} &
    \end{tabular}
    \caption{\textbf{Improving classification accuracy (of never-before-seen image classes) of a 3-T GOD
    subject, using its enhanced encoder via other subject:}
    The classification process, involves using an encoder to convert images without any fMRI measurements into fMRI patterns, even for semantic classes that the subject has never seen before. For each class, we feed 100 different images into the encoder, generating 100 corresponding fMRI patterns. These patterns are then averaged for each class, creating an ``average fMRI class representation" for each category. Consequently, when given a new test fMRI pattern, we can classify it by comparing it to these averaged fMRI class representations, to determine the correct class.
    }
    \label{S7}
\end{figure*}

\begin{figure*}[!hbt]
    \centering
    \begin{subfigure}{\textwidth}
        \includegraphics[width=\textwidth]{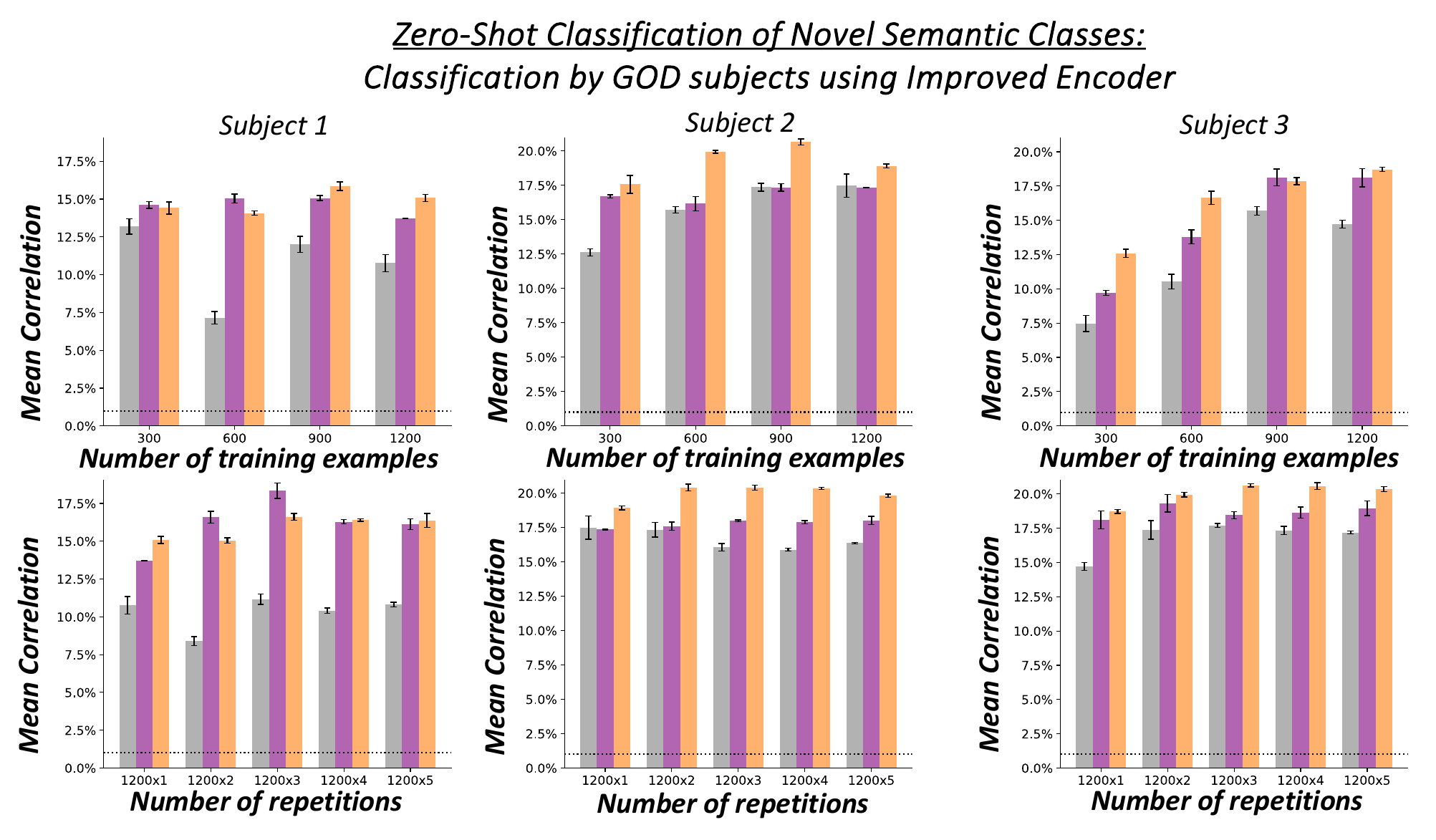} 
    \end{subfigure}
    \begin{subfigure}{\textwidth}
        \includegraphics[width=\textwidth]{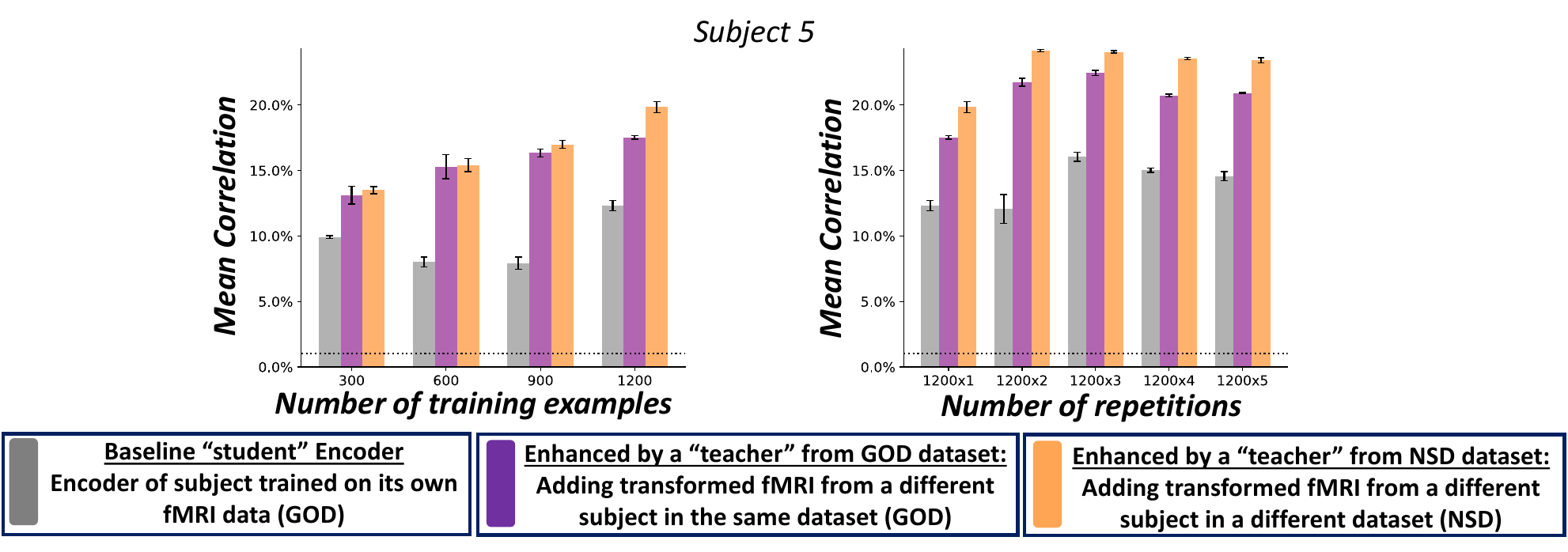} 
    \end{subfigure}
    
    \caption{\textbf{Image Classification from fMRI of 3-T GOD subjects:}
    This plot presents the classification results employing this approach, considering different modes of encoder training. We compare the baseline encoder (trained only on its subject-specific fMRI data) with encoders improved by another subject (``teacher"), either from the GOD dataset or using the NSD 7-Tesla dataset. The results presented are the mean accuracy of all subjects with the SEM error over different subjects and individual subject result with mean and SEM over 3 training repetitions.}
    \label{SM_Figure:Classification}
\end{figure*}

\begin{table}[ht!]
\centering
\begin{tabular}{|c|c|c|c|c|c|c|c|c|}
\hline
\multirow{2}{*}{\textbf{(Shared, Non-Shared)}} & \multicolumn{8}{c|}{\textbf{Subject}} \\ \cline{2-9} 
 & \textbf{1} & \textbf{2} & \textbf{3} & \textbf{4} & \textbf{5} & \textbf{6} & \textbf{7} & \textbf{8} \\ \hline
(100, 0) & 0.2124 & 0.2484 & 0.1424 & 0.1513 & 0.2274 & 0.1449 & 0.1130 & 0.1052 \\ \hline
(200, 0) & 0.2745 & 0.3094 & 0.1766 & 0.1954 & 0.2827 & 0.1882 & 0.1532 & 0.1131 \\ \hline
(400, 0) & 0.3204 & 0.3585 & 0.2226 & 0.2302 & 0.3213 & 0.2207 & 0.2018 & 0.1368 \\ \hline
(700, 0) & 0.3608 & 0.3902 & 0.2542 & 02514 & 0.3651 & 0.2487 & 0.2362 & 0.1625 \\ \hline
(700, 900) & 0.4272 & 0.4591 & 0.3172 & 0.3272 & 0.4334 & 0.3278 & 0.3042 & 0.2197 \\ \hline
(700, 2500) & 0.4581 & 0.4915 & 0.3615 & 0.3672 & 0.4681 & 0.3685 & 0.3513 & 0.2529 \\ \hline
(700, 5700) & 0.4753 & 0.5033 & 0.3819 & 0.3818 & 0.4837 & 0.3880 & 0.3651 & 0.2775 \\ \hline
(700, 6400) & 0.4752 & 0.5040 & 0.3761 & 0.3862 & 0.4796 & 0.3862 & 0.3677 & 0.2808 \\ \hline
\end{tabular}
\caption{\textbf{NSD encoders correlation:} Mean correlation values for each subject and example number pair (shared, non-shared) for all subjects in the NSD dataset.}
\label{table:mean_correlations}
\end{table}

\end{document}